\documentclass{article}
\usepackage{epsfig}

\begin{document}

\title{\Large \bf The azimuthal decorrelation of jets widely separated in rapidity as a test of the BFKL kernel}
\author{Agust{\' \i}n~Sabio~Vera$^1$ and Florian~Schwennsen$^2$\\[2.5ex]
$^1$ {\it Physics Department, Theory Division, CERN,}\\{\it CH--1211, Geneva 23, Switzerland}\\
$^2$ {\it II. Institut f\"{u}r Theoretische Physik, Universit\"{a}t Hamburg,}\\
 {\it Luruper Chaussee 149, D--22761 Hamburg, Germany}}

\maketitle

\vspace{-9cm}
\begin{flushright}
{\small CERN--PH--TH/2007--029\\DESY--07--017}
\end{flushright}

\vspace{7cm}
\begin{abstract}
We study the decorrelation in azimuthal angle of Mueller--Navelet jets at 
hadron colliders within the BFKL formalism. We introduce NLO terms in the 
evolution kernel and present a collinearly--improved version of it for all 
conformal spins. We show how this further resummation has good convergence 
properties and is closer to the Tevatron data than a simple LO treatment. 
However, we are still far from a good fit. We offer estimates of these 
decorrelations for larger rapidity differences which should favor the onset 
of BFKL effects and encourage experimental studies of this observable at the 
LHC.
\end{abstract}

\section{Introduction}
\label{sec:Introduction}

In this paper we continue the analytic study of the azimuthal decorrelations 
in Mueller--Navelet jets using 
the Balistky--Fadin--Kuraev--Lipatov (BFKL) equation~\cite{FKL} beyond the 
leading order 
approximation initiated in Ref.~\cite{Vera:2006un}. We investigate the 
inclusive hadroproduction of a pair of jets with large and similar transverse 
momenta produced at a large relative rapidity separation 
Y~\cite{Mueller:1986ey}. In principle, when 
this rapidity Y between the most forward and most backward jets is small there 
is not much phase space for the production of extra radiation and a fixed order 
perturbative calculation should be enough to describe the observable. However, 
if Y is large enough to make the product ${\alpha}_s Y \sim 1$ then a BFKL 
resummation of these terms to all orders should improve the accuracy of the 
calculation.

In Section~\ref{sec:Crosssections} we define our observables at partonic 
level and argue that the particular choice of rapidity variable removes the 
dependence on the parton distribution functions in normalized cross sections. 
The explicit expressions for angular differential cross sections including 
the next--to--leading (NLO) BFKL kernel~\cite{FLCC} with full angular 
dependence are also 
presented. In Section~\ref{sec:Resummation} we study in detail the structure 
of the scale invariant NLO BFKL kernel for different conformal spins. We show 
how the convergence of this kernel in terms of asymptotic intercepts 
is poor for zero conformal spin while being much better for larger ones. After 
this, a 
prescription to improve the collinear structure of the kernel using a shifted 
anomalous dimension for the full angular dependence is derived. 
The main consequence of this study is that even though collinear poles are 
removed for all conformal spins only the 
asymptotic intercept corresponding to 
the angular averaged case is modified, while the intercepts of the angular 
dependent components are hardly affected by the resummation. In 
Section~\ref{sec:Phenomenology} we use these kernels to make predictions 
for hadron colliders. We further discuss the need of a collinear 
resummation to generate stable results against a change of renormalization 
scheme and show that the resummed results provide a better description of 
the Tevatron data than those using LO BFKL. However, our results provide too 
much decorrelation when they are compared to the experimental measurements. 
From 
the theoretical side this could be due to the approximations needed to obtain 
our analytic results, {\it e.g.}, the  choice of rapidity variable and use of 
leading order jet vertices. In principle, our predictions should be more 
reliable at larger rapidities where multiple emissions between the two tagged 
jets are favored. It would be very interesting to further study these 
Mueller--Navelet jets at the Large Hadron Collider (LHC) at CERN to gauge the 
importance of configurations in multi--Regge and quasi--multi--Regge 
kinematics and try to investigate other possible jet topologies dominated 
by them. This point is further discussed at the end of our work, in the 
Conclusions section.

\section{Cross sections} 
\label{sec:Crosssections}

\begin{figure}[htbp]
\vspace{2cm}
  \centering
  \includegraphics[width=8cm]{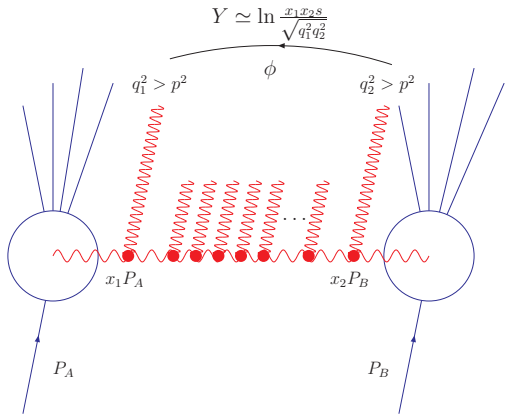}
  \caption{Representation of Mueller--Navelet jets at a hadron collider.}
  \label{MNgraph}
\end{figure}

We are interested in the study of normalized differential cross sections which 
turn out to be quite insensitive to the parton distribution functions. 
Therefore, to a good accuracy, the present analysis can be performed at 
partonic level and we focus on the 
${\rm parton} + {\rm parton} \rightarrow {\rm jet} + {\rm jet} + {\rm soft} 
\, \, {\rm emission}$ process. The differential cross section is then 
\begin{eqnarray}
\frac{d {\hat \sigma}}{d^2\vec{q}_1 d^2\vec{q}_2} &=& \frac{\pi^2 {\bar \alpha}_s^2}{2} 
\frac{f \left(\vec{q}_1,\vec{q}_2,{\rm Y}\right)}{q_1^2 q_2^2},
\end{eqnarray}
where ${\bar \alpha}_s = \alpha_s N_c / \pi$ is the strong coupling, 
$\vec{q}_{1,2}$ are the transverse momenta of the tagged jets, and Y their 
relative rapidity. The influence of the distribution functions is larger if 
the exact definition of the rapidity difference as in Fig.~\ref{MNgraph} is 
considered. For convenience we take Y as a fixed parameter since this allows 
us to perform all the necessary Mellin transforms and proceed with our 
analysis analytically. In more detail, the choice 
${\rm Y}=\ln{(x_1 x_2 s /s_0)}$ corresponds to a change of the energy scale 
$s_0=\sqrt{\vec{q}_1^2\vec{q}_2^2}$ to a fixed value. This change can be also 
understood as a NLO contribution to the vertices coupling the tagged jets to 
the external hadrons. The uncertainty associated to the choice of $s_0$ will 
be taken into account in the phenomenological discussion at the end of 
Section \ref{sec:Phenomenology}.

The gluon Green's function 
\begin{eqnarray}
f \left(\vec{q}_1,\vec{q}_2,{\rm Y}\right) &=& \int \frac{d\omega}{2 \pi i} e^{\omega {\rm Y}} f_\omega \left(\vec{q}_1,\vec{q}_2\right),
\end{eqnarray}
carries the full dependence on Y and fulfills the NLO BFKL equation which, in 
the transverse momenta operator representation
\begin{eqnarray}
{\hat q} \left|\vec{q}_i\right> &=& \vec{q}_i \left|\vec{q}_i\right>, 
\end{eqnarray}
with normalization
\begin{eqnarray}
\left<\vec{q}_1\right|{\hat 1}\left|\vec{q}_2\right> &=& \delta^{(2)} \left(\vec{q}_1-\vec{q}_2\right), 
\end{eqnarray}
can be written as
\begin{eqnarray}
\left(\omega - {\bar \alpha}_s {\hat K}_0 - {\bar \alpha}_s^2 {\hat K}_1\right) {\hat f}_\omega &=& {\hat 1}. 
\end{eqnarray}
This representation is useful when it acts on the basis
\begin{eqnarray}
\left< \vec{q}\right|\left.\nu,n\right> &=& \frac{1}{\pi \sqrt{2}} 
\left(q^2\right)^{i \nu -\frac{1}{2}} \, e^{i n \theta}, 
\label{eignfns}
\end{eqnarray}
which includes the dependence on the modulus and azimuthal angle of the 
different emissions. The Mellin--conjugate variable of the modulus is the 
real parameter $\nu$, and the Fourier--conjugate  parameter of the 
angle is the integer conformal spin $n$.

As the rapidity difference increases the azimuthal angle dependence is mainly 
driven by the kernel. This is the reason why, in the present work, we make use 
of the LO jet vertices which are much simpler than the NLO ones calculated in 
Ref.~\cite{impactfactors}. Following Ref.~\cite{Vera:2006un} we can then write 
the differential cross section in the azimuthal angle $\phi=\theta_1-\theta_2
-\pi$, where $\theta_i$ are the angles corresponding to the two tagged jets, as
\begin{eqnarray}
\frac{d {\hat \sigma}\left(\alpha_s, {\rm Y},p_{1,2}^2\right)}{d \phi}  &=&
\frac{\pi^2 {\bar \alpha}_s^2}{4 \sqrt{p_1^2 p_2^2}} \sum_{n=-\infty}^\infty 
e^{i n \phi} \, {\cal C}_n \left({\rm Y}\right),
\label{differential1}
\end{eqnarray}
with
\begin{eqnarray}
{\cal C}_n \left({\rm Y}\right) &=&
\frac{1}{2 \pi}\int_{-\infty}^\infty \frac{d \nu}{\left(\frac{1}{4}+\nu^2\right)}\left(\frac{p_1^2}{p_2^2}\right)^{i \nu} e^{\chi \left(\left|n\right|,\frac{1}{2}+ i \nu,{\bar \alpha}_s \left(p_1 p_2\right)\right){\rm Y} },
\label{Cn}
\end{eqnarray}
and
\begin{eqnarray}
\chi \left(n,\gamma,{\bar \alpha}_s\right) &\equiv&
{\bar \alpha}_s \chi_0\left(n,\gamma\right)
+{\bar \alpha}_s^2 \left(\chi_1\left(n,\gamma\right)
-\frac{\beta_0}{8 N_c} \frac{\chi_0\left(n,\gamma\right)}
{\gamma \left(1-\gamma\right)}\right).
\label{chigeneral}
\end{eqnarray}
Throughout this work, the coefficients ${\cal C}_n$ are not evaluated at the saddle point, but obtained by a  numerical integration over the full range of $\nu$. In the above expression the LO kernel, $\hat{K}_0$, has as eigenvalue the function
\begin{eqnarray}
\chi_0 \left(n,\gamma\right) &=& 2 \psi \left(1\right) - \psi \left(\gamma+ \frac{n}{2}\right) - \psi\left(1-\gamma+\frac{n}{2}\right), 
\end{eqnarray}
with $\psi$ being the logarithmic derivative of the Euler gamma function. 
The last term in Eq.~(\ref{chigeneral}) stems from the scale dependent part of the NLO kernel, {\it i.e.} from the running of the coupling. Its explicit form, in our representation,  depends on the impact factors and is discussed in more detail in Refs.~\cite{Vera:2006un,FSthesis}.
The action 
of the scale invariant sector of the NLO correction, $\hat{K}_1$, in the 
$\overline{\rm MS}$ renormalization scheme, on the 
basis in Eq.~(\ref{eignfns}) explicitly reads~\cite{Kotikov:2000pm}
\begin{eqnarray}
\chi_1\left(n,\gamma \right) &=& {\cal S} \chi_0 \left(n, \gamma\right)
+ \frac{3}{2} \zeta\left(3\right) - \frac{\beta_0}{8 N_c}\chi_0^2\left(n,\gamma\right)\nonumber\\
&+&\frac{1}{4}\left[\psi''\left(\gamma+\frac{n}{2}\right)+\psi''\left(1-\gamma+\frac{n}{2}\right)-2 \,\phi\left(n,\gamma\right)-2 \,\phi\left(n,1-\gamma\right)\right]\nonumber\\
&-&\frac{\pi^2 \cos{\left(\pi \gamma\right)}}{4 \sin^2\left(\pi \gamma\right)\left(1-2\gamma\right)}\left\{\left[3+\left(1+\frac{n_f}{N_c^3}\right)\frac{2+3\gamma\left(1-\gamma\right)}{\left(3-2\gamma\right)\left(1+2\gamma\right)}\right]\delta_n^0\right.\nonumber\\
&&\left.\hspace{2cm}-\left(1+\frac{n_f}{N_c^3}\right)\frac{\gamma\left(1-\gamma\right)}{2\left(3-2\gamma\right)\left(1+2\gamma\right)}\delta_n^2\right\},
\label{chi1}
\end{eqnarray}
where we have used the notation ${\cal S} = \left(4-\pi^2+5 \beta_0/N_c\right)/12$, $\beta_0 = (11 N_c-2 n_f)/3$ and $\zeta (n) = \sum_{p=1}^\infty p^{-n}$ is 
the Riemann zeta function. The $\phi (n,\gamma)$ function is 
of the form 
\begin{eqnarray}
\phi(n,\gamma) &=& \sum_{k=0}^{\infty} 
\frac{(-1)^{(k+1)}}{k+\gamma+\frac{n}{2}}
\left(\frac{}{}\psi'(k+n+1)-\psi'(k+1)\right.\nonumber\\
&&\left.\hspace{-2cm}+(-1)^{(k+1)} \left(\beta'(k+n+1)+\beta'(k+1)\right)+\frac{\psi(k+1)-\psi(k+n+1)}{k+\gamma+\frac{n}{2}}\right),
\end{eqnarray}
with
\begin{eqnarray}
4 \,\beta'(\gamma) &=& \psi'\left(\frac{1+\gamma}{2}\right)
-\psi'\left(\frac{\gamma}{2}\right).
\end{eqnarray}
These scale invariant eigenvalues have a very interesting structure which 
we will study in the next section. Before this we would like to indicate 
that the full cross section corresponds to the integration over the azimuthal 
angle of the differential expression in Eq.~(\ref{differential1}). This 
implies that it only depends on the $n=0$ component:
\begin{eqnarray}
{\hat \sigma}\left(\alpha_s, {\rm Y},p_{1,2}^2\right) &=& 
\frac{\pi^3 {\bar \alpha}_s^2}{2 \sqrt{p_1^2 p_2^2}} \, {\cal C}_0 \left({\rm Y}\right).
\end{eqnarray}
In this paper we are interested in those distributions which are sensitive to 
the higher conformal spins. In particular, the average of the cosine of the 
azimuthal angle times an integer projects out the contribution from each of 
these angular components. It can be obtained using the ratio
\begin{eqnarray}
\left<\cos{\left( m \phi \right)}\right> &=& \frac{{\cal C}_m \left({\rm Y}\right)}{{\cal C}_0\left({\rm Y}\right)}.
\label{cosmphi}
\end{eqnarray}
The associated ratios
\begin{eqnarray}
\frac{\left<\cos{\left( m \phi \right)}\right>}{\left<\cos{\left( n \phi \right)}\right>} &=& \frac{{\cal C}_m \left({\rm Y}\right)}{{\cal C}_n\left({\rm Y}\right)}
\label{Ratiosformula}
\end{eqnarray}
are also of interest since they can be used to remove the uncertainty 
associated to the hard pomeron intercept, {\it i.e.} the $n=0$ component, 
which we will analyze below. To study the behavior of all the angular 
components together it is useful to use the normalized differential cross 
section on the azimuthal angle:
\begin{eqnarray}
\frac{1}{{\hat \sigma}}\frac{d{\hat \sigma}}{d \phi} ~=~
\frac{1}{2 \pi}\sum_{n=-\infty}^\infty 
e^{i n \phi} 
\frac{{\cal C}_n\left({\rm Y}\right)}
     {{\cal C}_0\left({\rm Y}\right)} 
~=~ \frac{1}{2\pi}
\left\{1+2 \sum_{n=1}^\infty \cos{\left(n \phi\right)}
\left<\cos{\left( n \phi \right)}\right>\right\}.
\label{fullangular}
\end{eqnarray}
Before comparing the results stemming from these expressions with the 
experimental data it is important to first analyze the convergence of the 
kernel at NLO for the different conformal spins. We proceed with this study 
in the coming section.

\section{Collinear resummation} 
\label{sec:Resummation}

It is well--known that the BFKL resummation presents an instability 
when the NLO corrections are taken into account, for details regarding this 
point, see, {\it e.g.}, Refs.~\cite{gavin1,improvedkernel,agus05}. 
For the observables studied in this paper we have found that if we use 
the NLO BFKL kernel 
as it stands the cross sections are very dependent on the renormalization 
scheme. 
In particular, the term proportional to $\chi_0$ in $\chi_1$ can be removed by  a shift of the Landau pole of the form
\begin{eqnarray}
\Lambda_{\overline{\rm MS}} \rightarrow \Lambda_{\rm GB} = \Lambda_{\overline{\rm MS}} 
e^{\frac{2 N_c}{\beta_0}{\cal S}}.
\end{eqnarray}
This defines the so--called gluon--bremsstrahlung (GB) 
scheme~\cite{Catani:1990rr,Dokshitzer:1995ev} which is commonly used when 
dealing with soft gluon resummations. In this new scheme we find that 
some of our distributions even change sign and become unphysical. This is a manifestation of 
the poor convergence of the series. A crucial ingredient to improve the convergence of the 
perturbative expansion is to 
demand compatibility of the BFKL kernel with renormalization group evolution 
to all orders in the limit of deep inelastic 
scattering. This can be achieved if corrections to all orders are 
introduced by means of a shift 
in the anomalous dimension~\cite{Ciafaloni:2003rd}. So far these types of resummations have 
been performed for a BFKL kernel averaged over the azimuthal angle and therefore they only 
affect the zero conformal spin sector. For our purposes in this work we must study the 
convergence of the eigenvalues of the kernel for all angular components. 

The renormalization group improved kernels are based on a proper treatment of 
the collinear region of emissions. This region is insensitive to the azimuthal 
angle so we should not find a big effect beyond the $n=0$ case when we resum. 
We will see that this is the case because the asymptotic 
intercepts for the different angular components with $n>0$ are very stable 
under radiative corrections. 

To start our investigation it is necessary to extract the pole structure of the different 
contributions to the kernel around the $\gamma = -\frac{n}{2}, 1+ \frac{n}{2}$ points. These 
poles are as follows:
\begin{eqnarray}
\chi_0 \left(n,\gamma\right) &\simeq& \frac{1}{\gamma +\frac{n}{2}} + 
\left\{\gamma \rightarrow 1 -\gamma \right\}, \\
\chi_1 \left(n,\gamma\right) &\simeq& \frac{a_n}{\gamma +\frac{n}{2}} + 
\frac{b_n}{\left(\gamma +\frac{n}{2}\right)^2} - 
\frac{1}{2\left(\gamma +\frac{n}{2}\right)^3} + 
\frac{c \,\delta_n^2}{\gamma} +
\left\{\gamma \rightarrow 1 -\gamma \right\}.
\end{eqnarray}
The coefficients at the singular points can be written as
\begin{eqnarray}
a_n &=& {\cal S} - \frac{\pi^2}{24}+\frac{\beta_0}{4 N_c} {\rm H}_n 
+\frac{1}{8}\left(\psi'\left(\frac{n+1}{2}\right)-
\psi'\left(\frac{n+2}{2}\right)\right) + \frac{1}{2} \psi'\left(n+1\right)
\nonumber\\
&-&\frac{\delta_n^0}{36} \left(67+13 \frac{n_f}{N_c^3}\right)
-\frac{47 \delta_n^2}{1800} \left(1+\frac{n_f}{N_c^3}\right),\\
- b_n &=& \frac{\beta_0}{8 N_c}+\frac{1}{2}{\rm H}_n
+\frac{\delta_n^0}{12} \left(11+2 \frac{n_f}{N_c^3}\right)
+\frac{\delta_n^2}{60} \left(1+\frac{n_f}{N_c^3}\right),\\
c &=& \frac{1}{24}\left(1+\frac{n_f}{N_c^3}\right).
\end{eqnarray}
Here ${\rm H}_n$ stands for the harmonic number 
$\psi\left(n+1\right)-\psi \left(1\right)$.

The term due to running coupling effects 
$-\frac{\beta_0}{8 N_c} \frac{\chi_0\left(n,\gamma\right)}{\gamma \left(1-\gamma\right)}$,  
in Eq.~(\ref{chigeneral}), introduces the following modification of the single and double 
NLO poles of the original kernel:
\begin{eqnarray}
a_n &\rightarrow& a_n + \frac{\beta_0}{2 N_c}
\left(\frac{1-\delta_n^0}{n(2+n)}-\frac{\delta_n^0}{4}\right),\\
b_n &\rightarrow& b_n - \frac{\beta_0}{8 N_c} \delta_n^0.
\end{eqnarray}

There is some freedom in the way the resummation of collinear terms can be 
performed. We find that the most natural scheme is an extension of that 
discussed in Ref.~\cite{agus05}, which was first proposed in~\cite{gavin1}, 
to include the 
dependence on all conformal spins. For completeness we have checked that the 
numerical results we will present are very similar for different resummations \cite{FSthesis}. In this way, to 
obtain a convergent series for all values of the conformal spins, we use the prescription
\begin{eqnarray}
\omega  &=& {\bar \alpha}_s \left(1+ {\cal A}_n {\bar \alpha}_s\right)
\left\{2 \, \psi \left(1\right) 
- \psi \left(\gamma + \frac{\left|n\right|}{2}+\frac{\omega}{2}+{\cal B}_n {\bar \alpha}_s \right) \right. \\
&&\hspace{-1cm}- \left.\psi \left(1-\gamma + \frac{\left|n\right|}{2}+\frac{\omega}{2}+{\cal B}_n {\bar \alpha}_s \right) \right\} + {\bar \alpha}_s^2 \, \Bigg\{\chi_1 \left(\left|n\right|,\gamma\right)-\frac{\beta_0}{8 N_c} \frac{\chi_0\left(n,\gamma\right)}{\gamma \left(1-\gamma\right)}\nonumber\\
&&\hspace{-1cm}-{\cal A}_n \chi_0\left(\left|n\right|,\gamma\right)\Bigg)
+ \left(\psi'\left(\gamma + \frac{\left|n\right|}{2}\right)
+\psi' \left(1-\gamma + \frac{\left|n\right|}{2}\right) \right)
\left(\frac{\chi_0\left(\left|n\right|,\gamma\right)}{2}+{\cal B}_n\right)
\Bigg\},\nonumber
\end{eqnarray}
where the ${\cal A}_n$ and ${\cal B}_n$ coefficients are related to those of the original NLO 
kernel by
\begin{eqnarray}
{\cal A}_n &=& a_n + \psi'\left(n+1\right),\\
{\cal B}_n &=& \frac{1}{2} {\rm H}_n - b_n.
\end{eqnarray}

It is worth noting that the solution to this transcendental equation can be approximated to a 
very good accuracy by the expression
\begin{eqnarray}
\label{approximation}
\omega &=& {\bar \alpha}_s \chi_0\left(\left|n\right|,\gamma\right) 
+ {\bar \alpha}_s^2 \Bigg(\chi_1\left(\left|n\right|,\gamma\right) -\frac{\beta_0}{8 N_c} \frac{\chi_0\left(n,\gamma\right)}{\gamma \left(1-\gamma\right)} 
\Bigg) \\
&&\hspace{-1.2cm}+ \Bigg\{ \sum_{m=0}^\infty \Bigg[-m + b_n \, {\bar \alpha}_s 
-\frac{\left|n\right|}{2}-\gamma + \sqrt{2 \left({\bar \alpha}_s + a_n \, {\bar \alpha}_s^2\right)+\left(m-b_n \, {\bar \alpha}_s +\gamma + \frac{\left|n\right|}{2}\right)^2} \nonumber\\
&&\hspace{-1.2cm}- \Bigg(
\frac{{\bar \alpha}_s + a_n  {\bar \alpha}_s^2}{\gamma + m + \frac{\left|n\right|}{2}} + 
\frac{{\bar \alpha}_s^2 \, b_n}{\left(\gamma + m +\frac{\left|n\right|}{2}\right)^2} -
\frac{{\bar \alpha}_s^2}{2\left(\gamma + m +\frac{\left|n\right|}{2}\right)^3} 
\Bigg)\Bigg]+\left\{\gamma \rightarrow 1 - \gamma\right\}\Bigg\}.\nonumber
\end{eqnarray}
This is an extension to the present case of the ``All--poles'' approximation 
developed in Ref.~\cite{agus05}. 

\begin{figure}[tbp]
\hspace{-.4cm}\epsfig{width=4.5cm,file=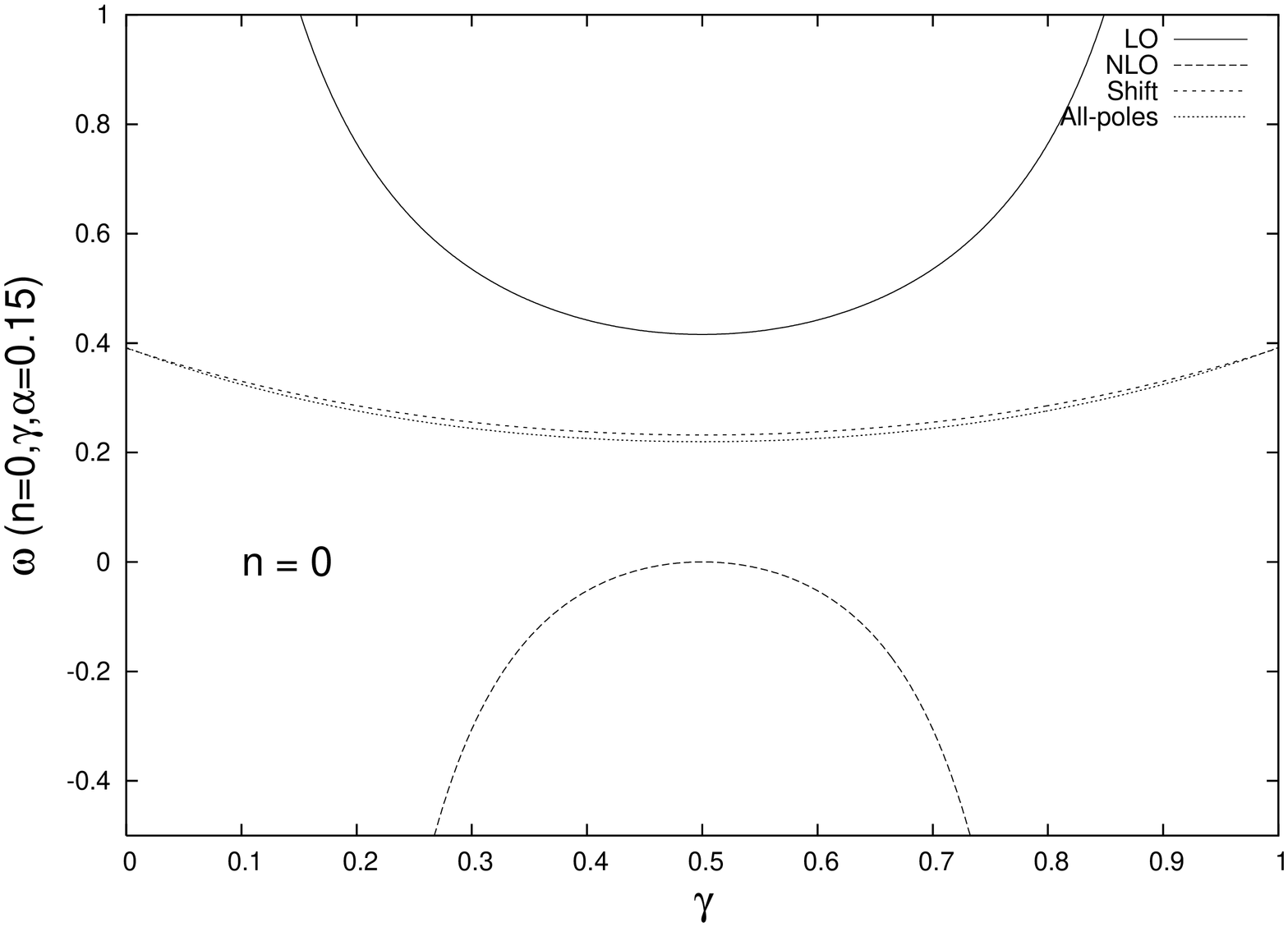,angle=-90}\epsfig{width=4.5cm,file=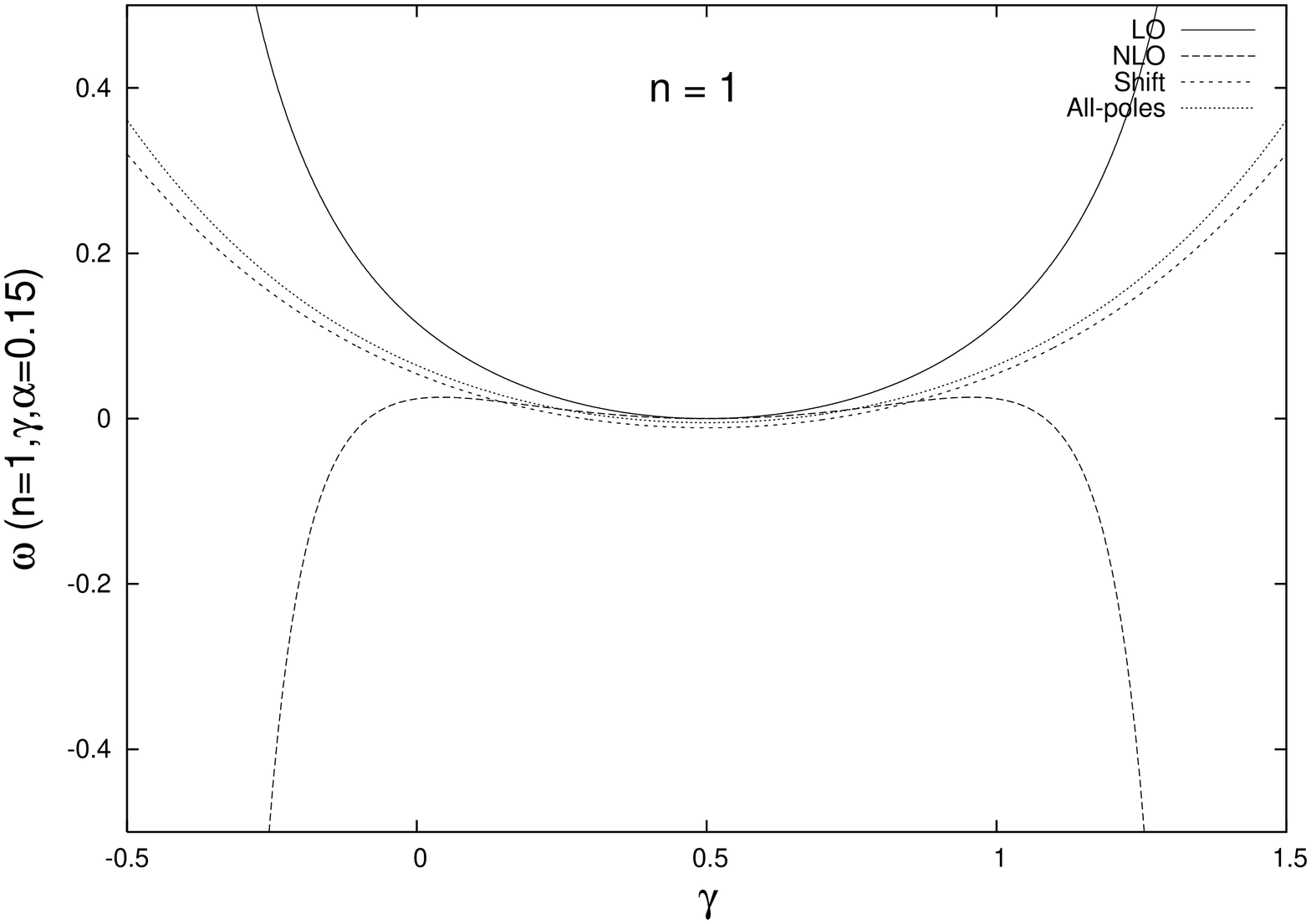,angle=-90}

\hspace{-.4cm}\epsfig{width=4.5cm,file=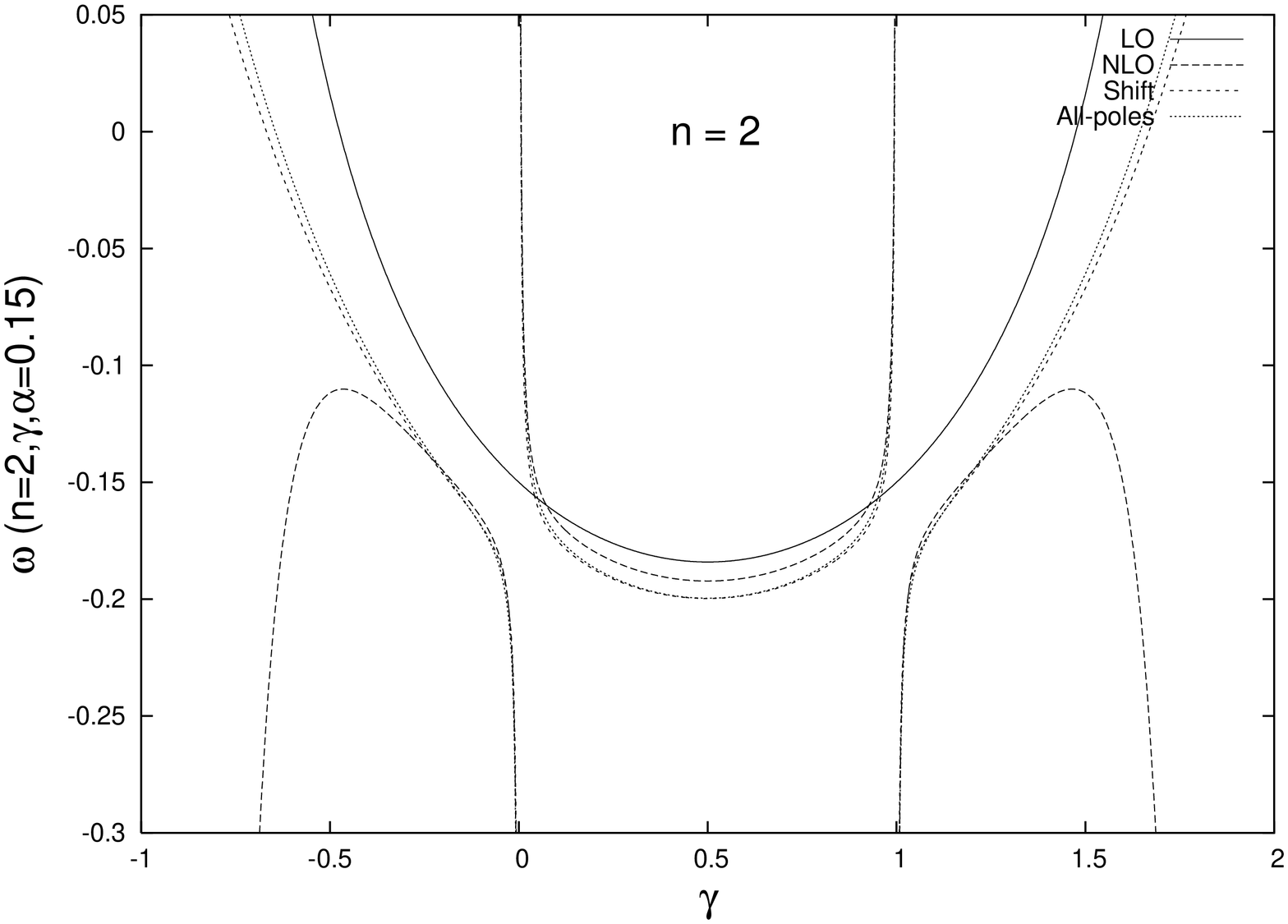,angle=-90}\epsfig{width=4.5cm,file=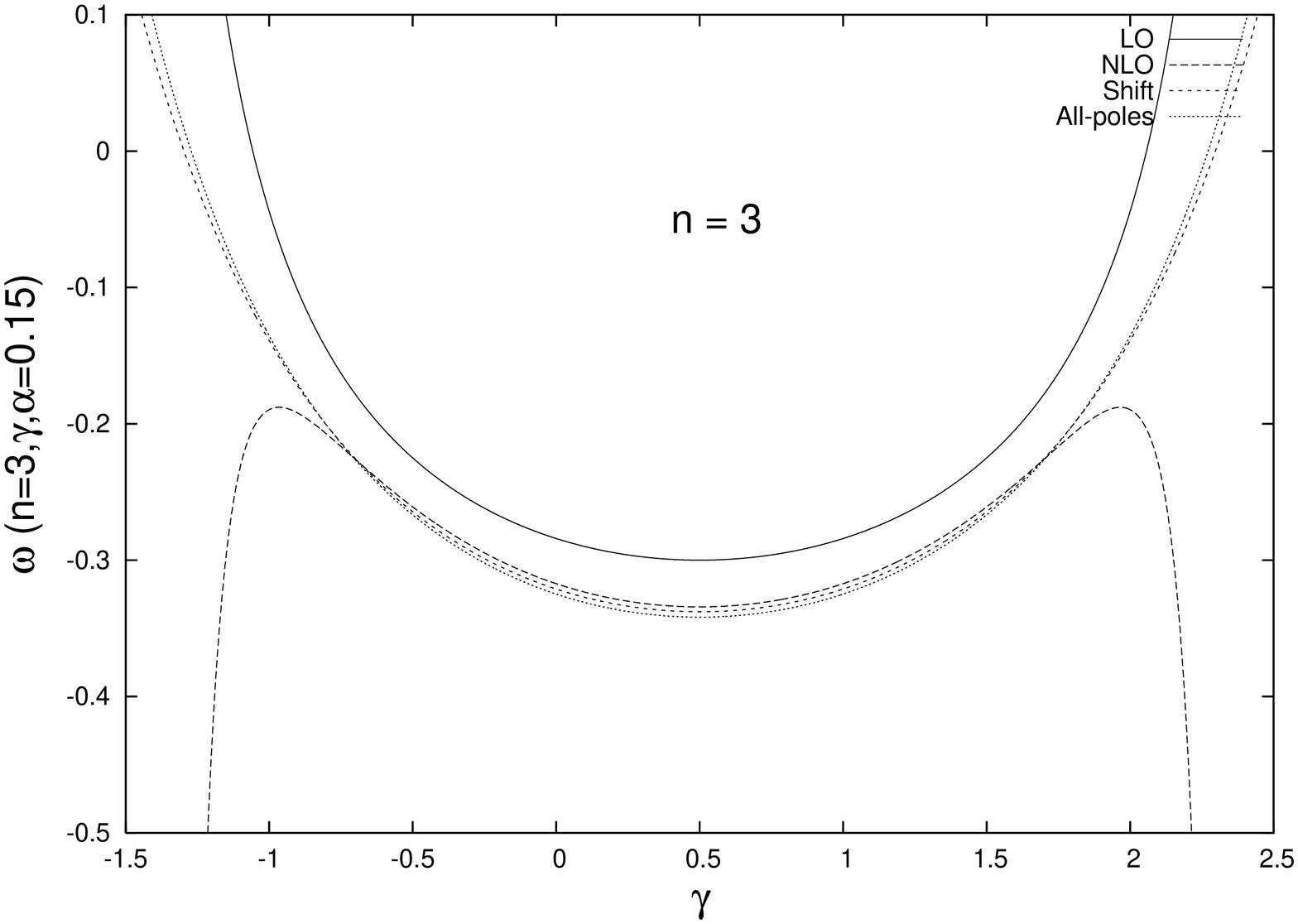,angle=-90}
\caption{Eigenvalues of the scale invariant sector of the BFKL kernel as 
a function of $\gamma$ for different values of the conformal spin.}
\label{kernelvsgamma}
\end{figure}

\begin{figure}[tbp]
\hspace{-.4cm}\epsfig{width=4.5cm,file=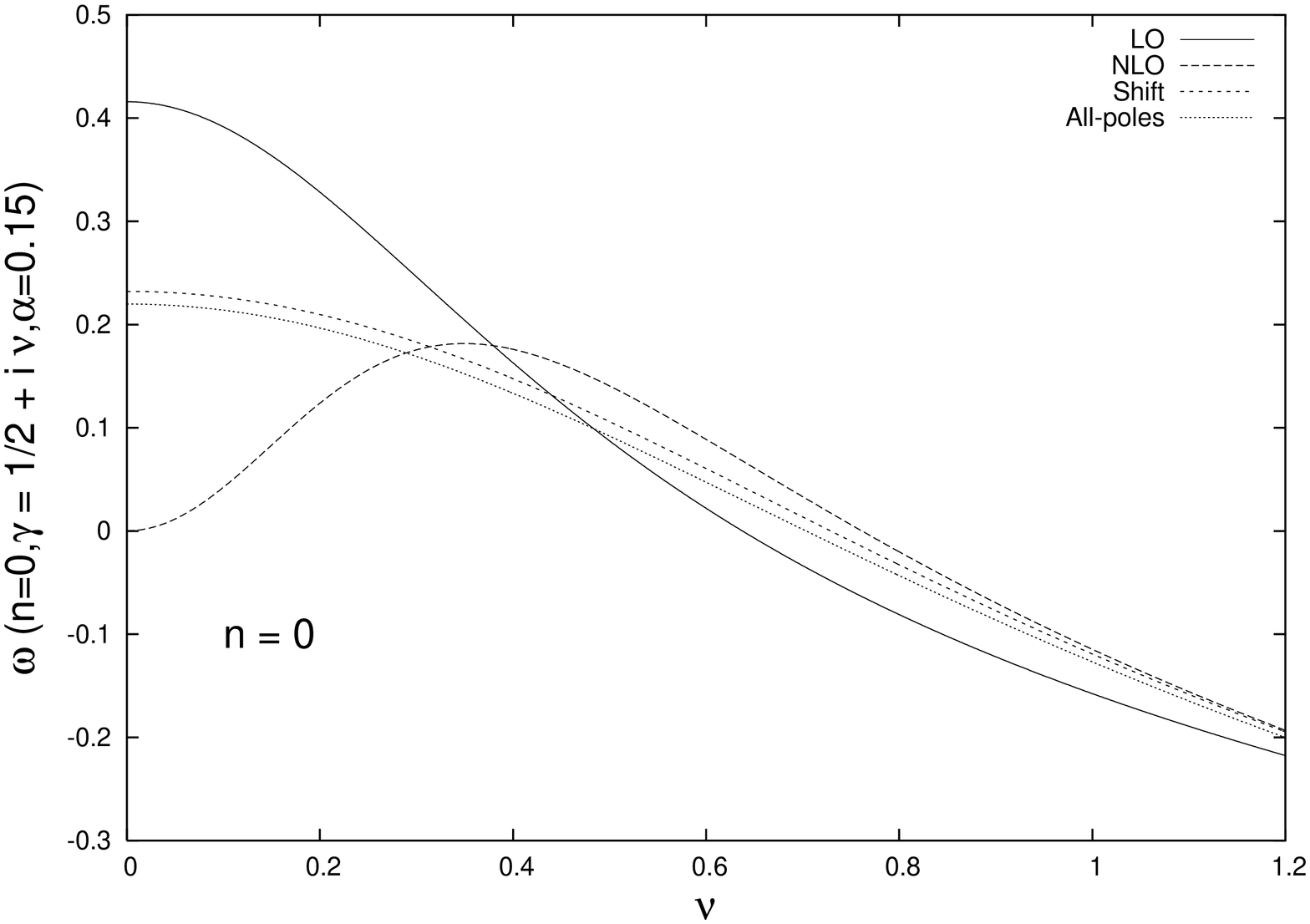,angle=-90}\epsfig{width=4.5cm,file=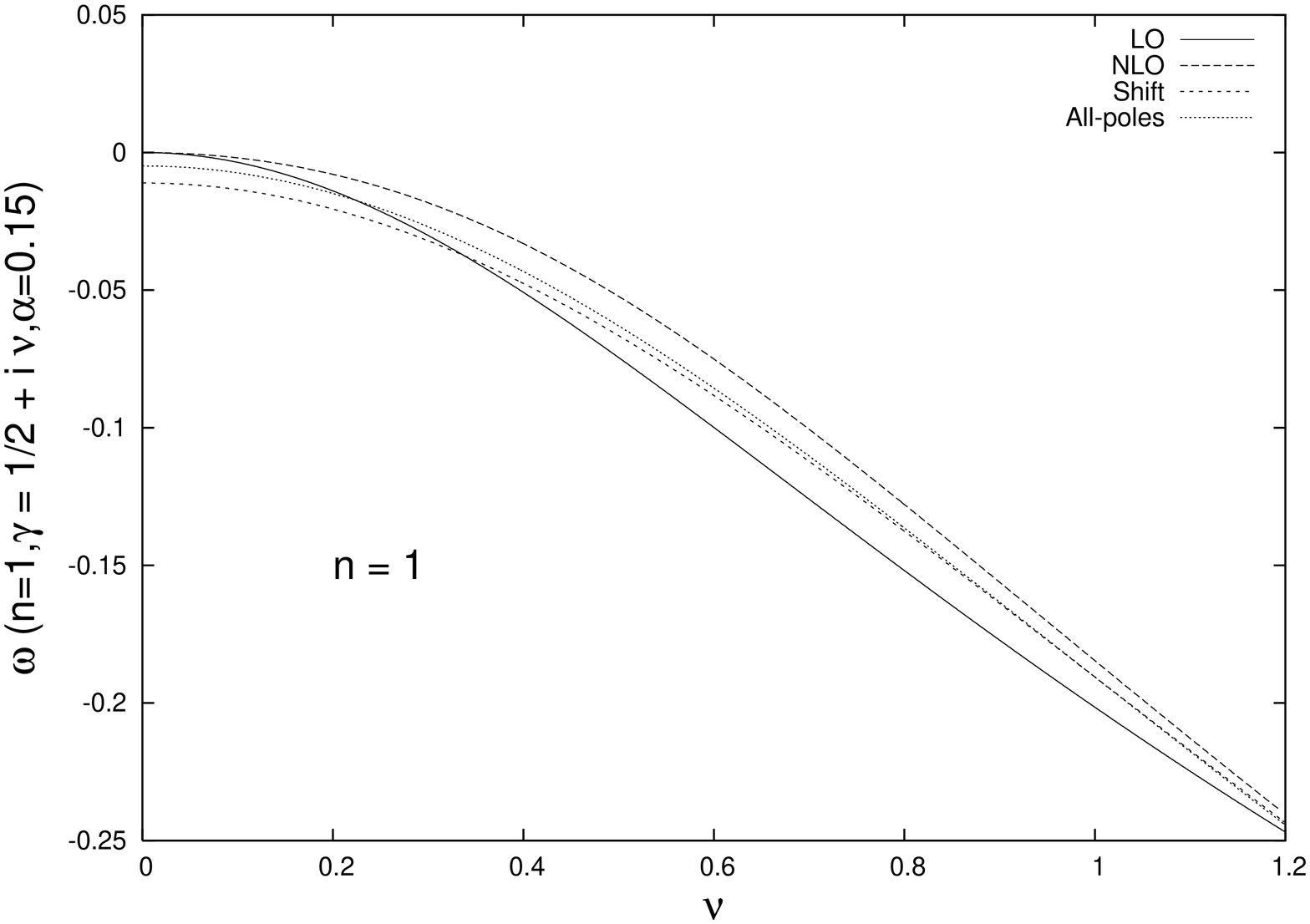,angle=-90}

\hspace{-.4cm}\epsfig{width=4.5cm,file=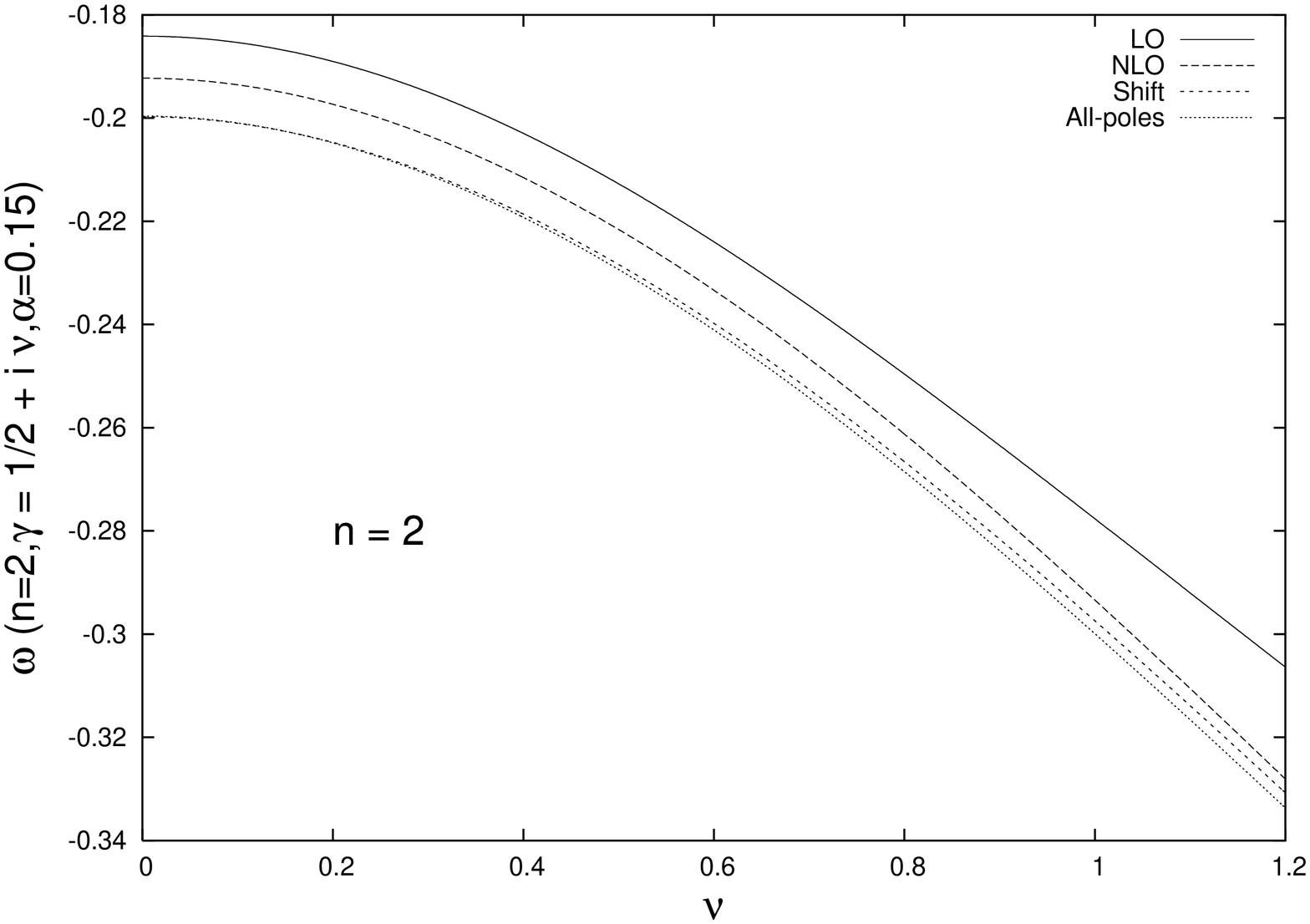,angle=-90}\epsfig{width=4.5cm,file=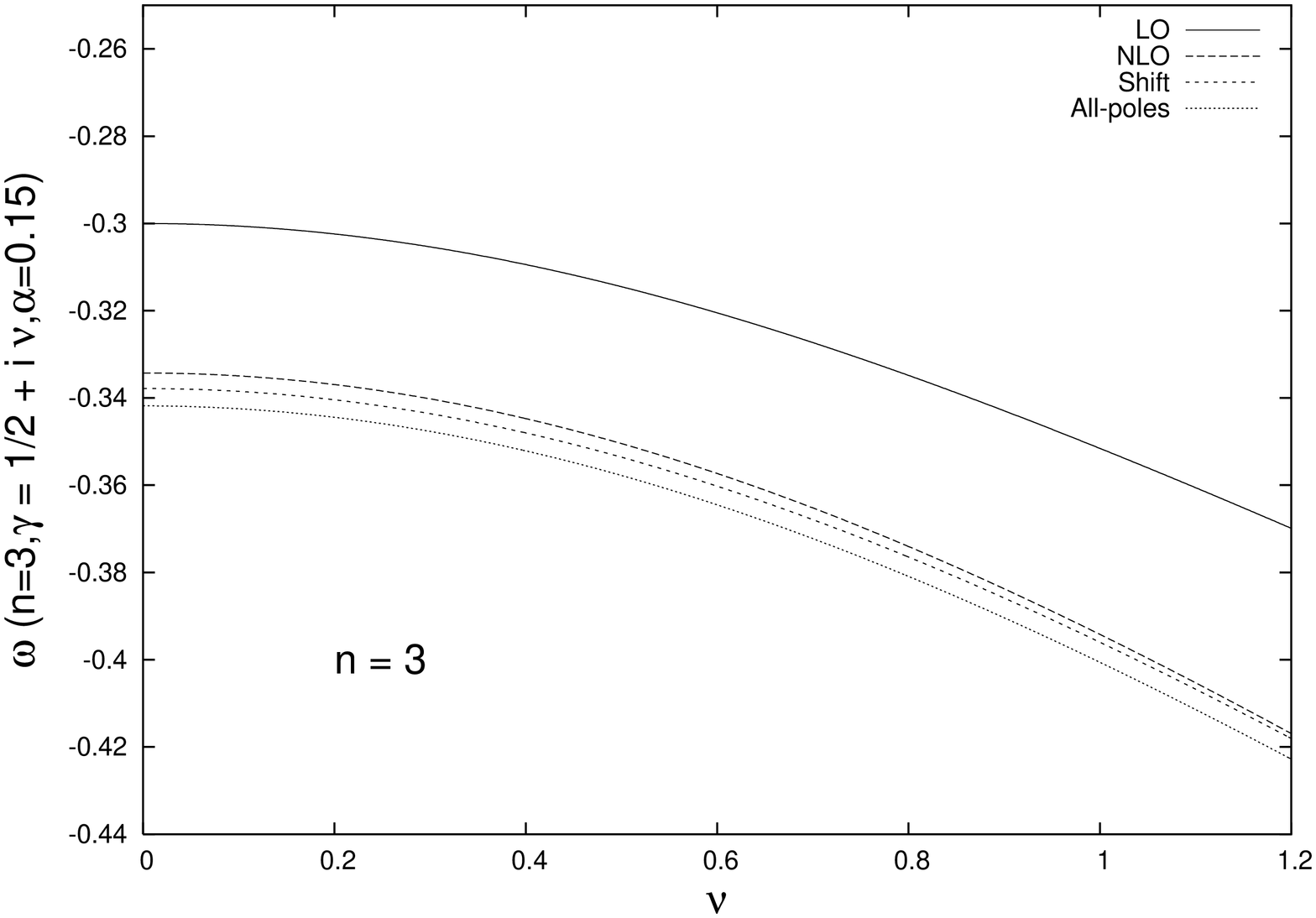,angle=-90}
\caption{Eigenvalues of the scale invariant sector of the BFKL kernel as 
a function of $\nu$ for different values of the conformal spin.}
\label{kernelvsnu}
\end{figure}

In the presentation of our resummed kernels and experimental observables we 
will be using the $\overline{\rm MS}$ renormalization scheme. We have checked that these 
results do not significantly change when the GB renormalization scheme is used 
instead. This gives us confidence on the stability of our calculations. In 
Fig.~\ref{kernelvsgamma} we have plotted the eigenvalue of the scale invariant 
sector of the BFKL kernel for different conformal spins. The value we have 
chosen for the coupling constant is ${\bar \alpha}_s = 0.15$. For simplicity, 
in these plots we have not included the term related to the LO impact factors. 
As a general feature we see how for non--zero conformal spins the LO and NLO 
kernels contain poles which are moving away from the $0<\gamma<1$ region. 
These singularities are removed when the collinear resummation, which is 
indicated as ``Shift'' in the plots, is introduced. We also show how the 
approximation provided by Eq.~(\ref{approximation}), denoted as 
``All--poles'', is a very good one. A remarkable feature of the NLO kernel 
takes place for conformal spin 2. Here the terms proportional to $\delta_n^2$ 
in Eq.~(\ref{chi1}) generate poles at $\gamma =0,1$ which we choose not to 
resum away. The structure for the higher $n$'s is the same as for $n=1,3$ with 
the poles ``traveling'' towards the left and right directions in $\gamma$. 
In these plots we can already see how the region relevant for asymptotic 
intercepts, around $\gamma \sim 0.5$, is only sensitive to the collinear 
resummation for the angular averaged component, $n=0$. For the other 
contributions the intercepts are practically invariant under the introduction 
of radiative corrections. To highlight this feature we study the region 
$\gamma = \frac{1}{2}+ i \nu$ for small $\nu$ in Fig.~\ref{kernelvsnu}. 
We recognize the familiar feature for $n=0$ where the double maxima are  
replaced by a single one at $\nu = 0$ when the matching to collinear evolution 
is performed, this removes unphysical oscillations in the gluon Green's 
function~\cite{agus05}. We also see that all the $n \neq 0$ contributions  
have a single maximum at that point and the resummation only shifts their  
asymptotic intercepts by a very small amount. 

\begin{figure}[tbp]
\hspace{-.4cm}\epsfig{width=4.5cm,file=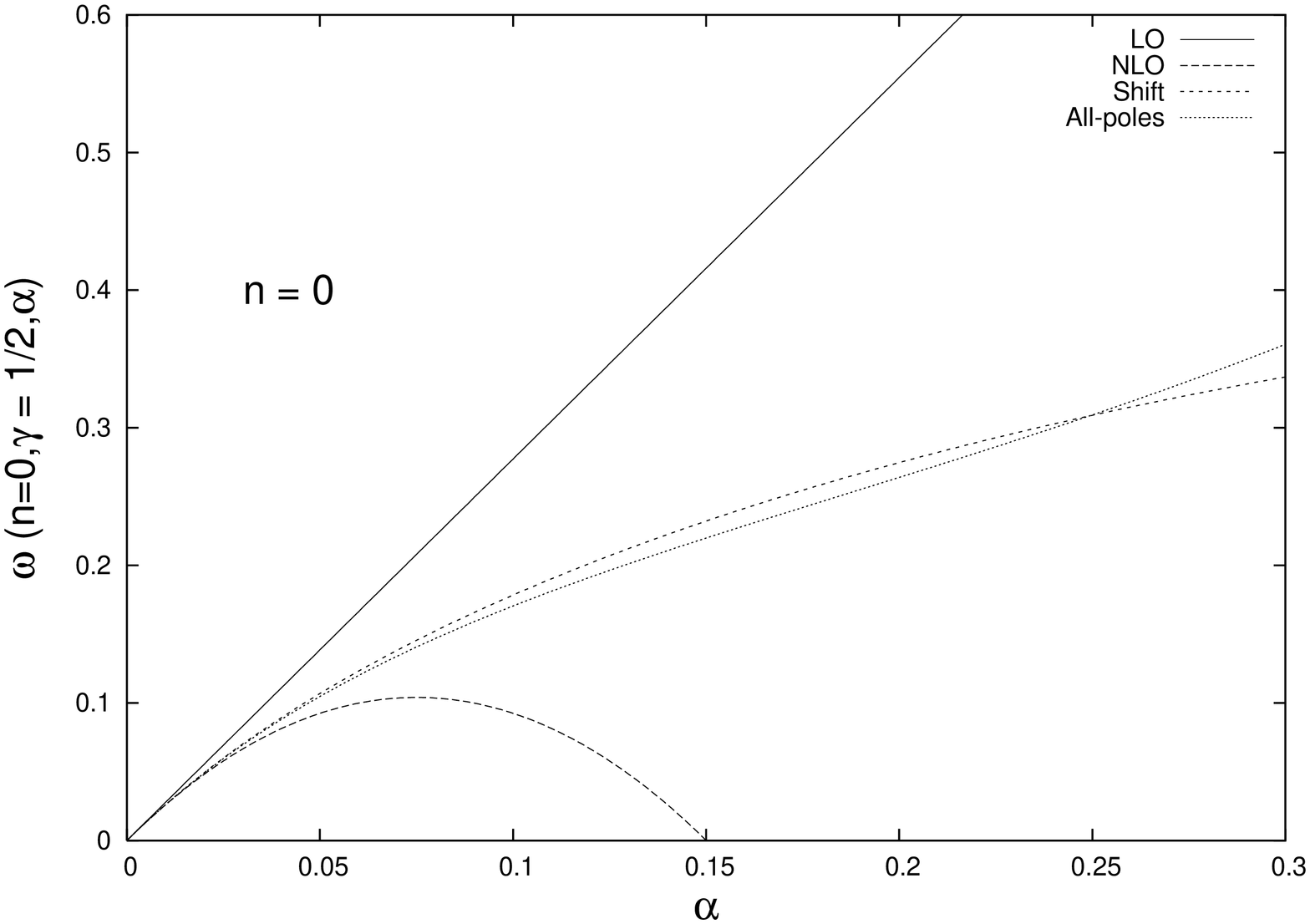,angle=-90}\epsfig{width=4.5cm,file=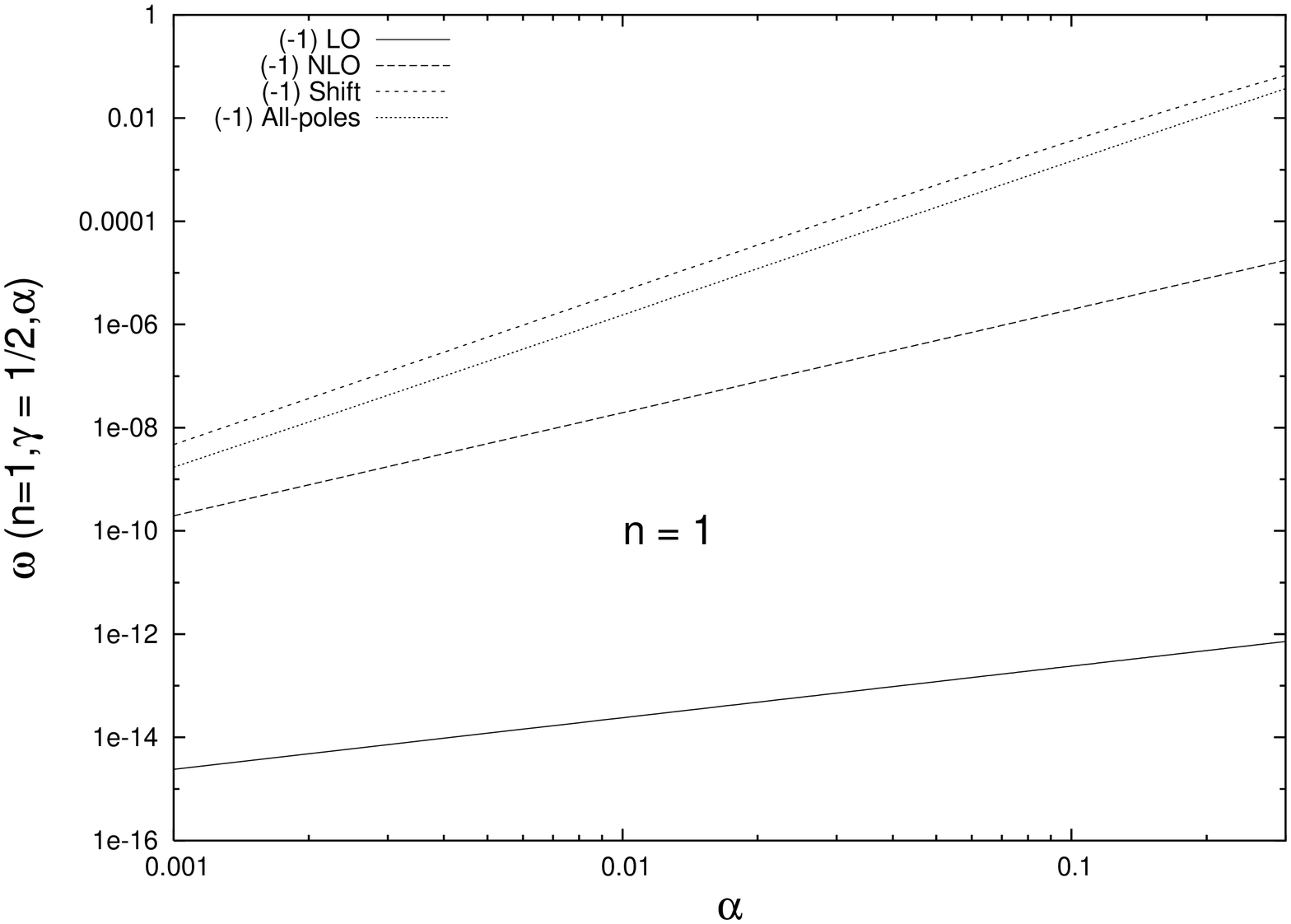,angle=-90}

\hspace{-.4cm}\epsfig{width=4.5cm,file=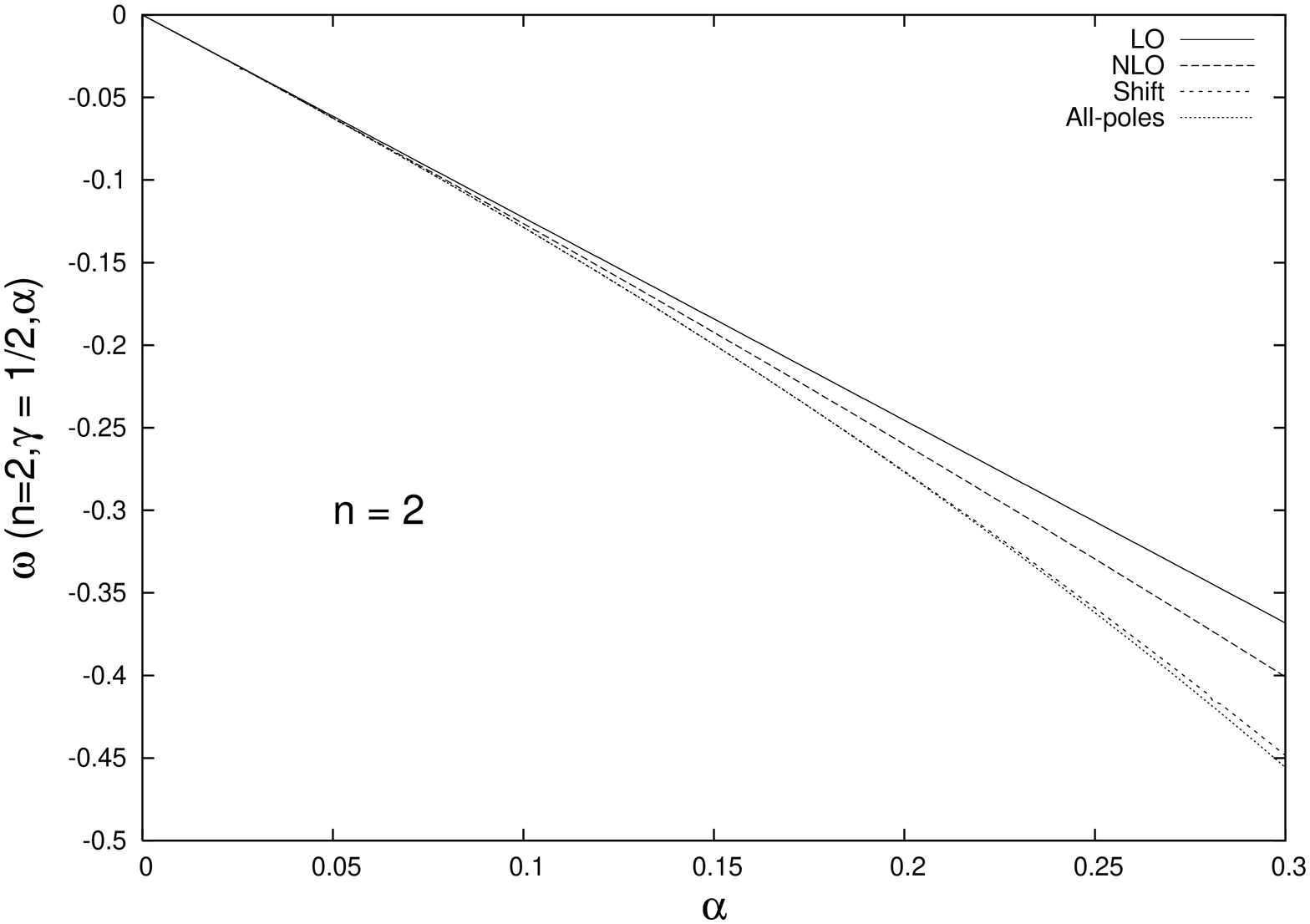,angle=-90}\epsfig{width=4.5cm,file=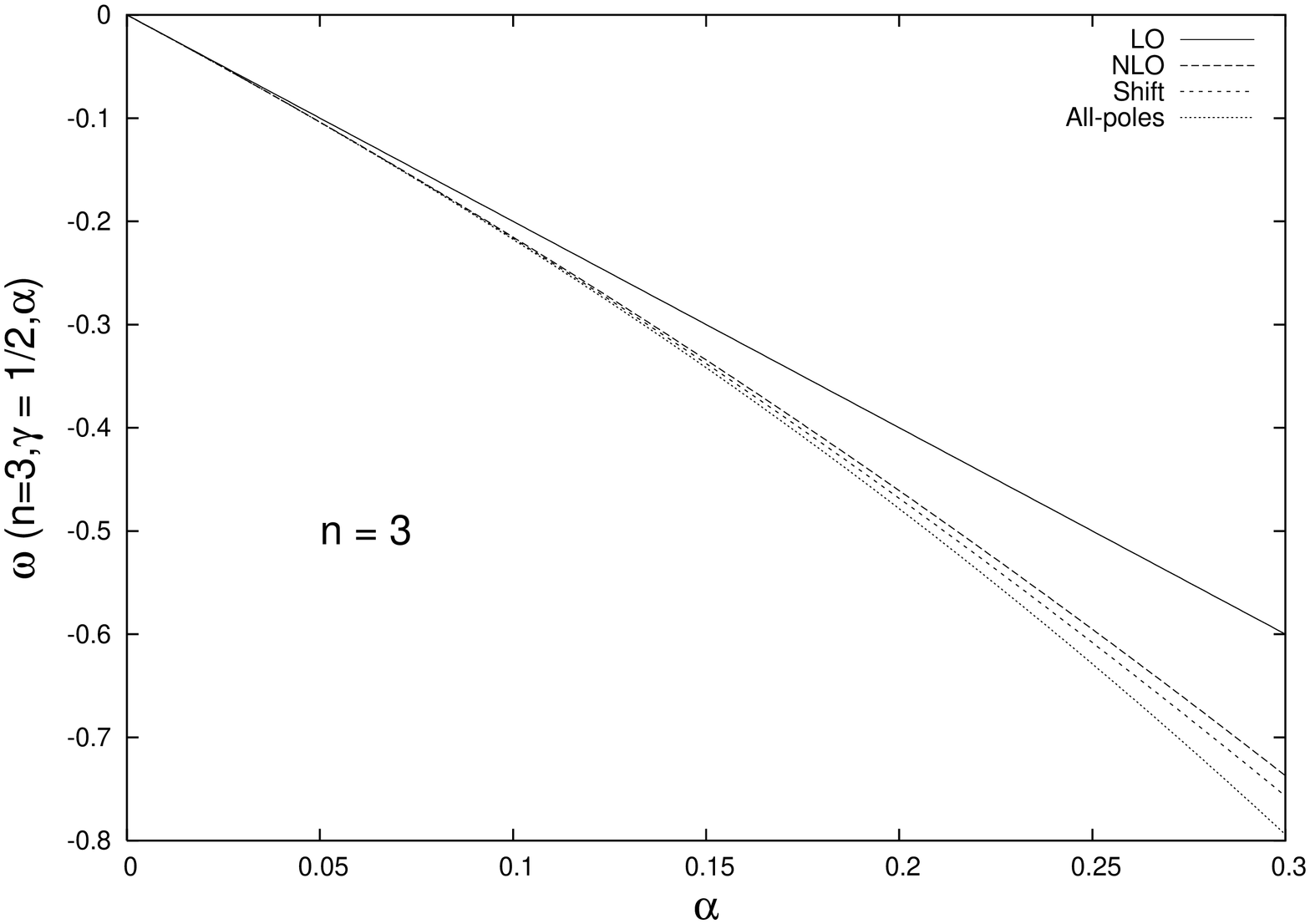,angle=-90}
\caption{Eigenvalues of the scale invariant sector of the BFKL kernel as a 
function of the coupling constant for different values of the conformal spin.}
\label{kernelvsalphas}
\end{figure}

We would like to point out that the only conformal spin with positive 
asymptotic intercept is $n=0$. 
This implies that the dependence on the azimuthal angles in the Green's 
function decreases with energy. This property remains when the value of 
the coupling increases as can be seen in Fig.~\ref{kernelvsalphas}.

After having introduced the resummed kernel used in our calculations we are 
now ready to compare with the experimental results at the Tevatron and make 
predictions for other colliders. This is done in Section~\ref{sec:Phenomenology}.

\section{Phenomenology} 
\label{sec:Phenomenology}

When dealing with Mueller--Navelet jets we are interested in hadron--hadron 
collisions where two jets are tagged in the very forward and very backward 
regions with similar semihard transverse momenta, $p^2$, such that $s \gg  
p^2 \gg \Lambda_{\rm QCD}^2$. For large rapidity differences between these 
two jets logarithms of the form $\left({\bar \alpha}_s \log{(s/p^2)}\right)^n$ 
should be 
resummed using the BFKL equation. Mueller and Navelet proposed this process 
in Ref.~\cite{Mueller:1986ey} as ideal to apply the BFKL formalism and 
predicted a power--like rise for the cross section. However, to realize this 
growth as a manifestation of multi--Regge kinematics is very difficult since 
it is drastically damped by the behavior of the parton distribution functions 
(PDFs) for $x \rightarrow 1$. A possible way out is to fix the PDFs and to 
vary the center--of--mass energy of the hadron collider itself, and thereby 
vary the rapidity difference, Y, between the two tagged jets. 
BFKL predicts a behavior of the cross sections of the form $\sigma \sim 
\exp{(\alpha - 1) Y}/\sqrt{Y}$ with $\alpha$ being the 
intercept. The D$\emptyset$ collaboration analyzed data taken at the Tevatron 
$p{\bar p}$--collider from two periods of measurement at different energies 
$\sqrt{s} = 630$ and 1800 GeV. From these they extracted an intercept 
of $1.65 \pm .07$~\cite{Abbott:1999ai}. This rise is even faster than that 
predicted in the LO BFKL calculation which for the kinematics relevant in the 
D$\emptyset$ experiment yields an approximated value of 1.45. It has been argued~\cite{MNDelDuca01} 
that the exact experimental and theoretical definitions of the cross sections 
disagreed making the interpretation of the results cumbersome, and the fact 
that the experimental determination of the intercept is based on just two 
data points leaves room for other possible explanations.

In this work we focus on a more exclusive observable, namely, the azimuthal 
angle decorrelation between the jets. We would like to recall that the 
Mueller--Navelet jets lie at the interface of collinear factorization and 
BFKL dynamics. The partons emitted from the hadrons carry large longitudinal 
momentum fractions and, after scattering off each other, they produce the two 
tagged jets. Because of the large transverse momentum of these jets, the 
partons are hard and obey collinear factorization. In particular, their scale 
dependence is governed by the DGLAP evolution equations. Between the jets, 
on the other hand, we require a large rapidity difference, a configuration 
largely dominated by multi--Regge kinematics. Therefore, the hadronic cross 
section factorizes into two conventional collinear parton distributions 
convoluted with the partonic cross section, described within the BFKL context. 
With respect to the partonic cross section, the incoming partons, 
consequently, are considered to be on--shell and collinear to the incident 
hadrons. 

For the angular correlation theoretical predictions from LO BFKL were first 
obtained in Refs.~\cite{DelDuca:1993mn,Stirling:1994zs}, improvements due to 
the running of the coupling and proper treatment of the kinematics have been 
implemented in Refs.~\cite{Orr:1997im,Kwiecinski:2001nh}. A first step 
towards an analytic NLO description has been made in 
Refs.~\cite{Vera:2006un,Vera:2006xa} on which our work builds up. Ten years 
ago the D$\emptyset$ collaboration at the Tevatron measured the azimuthal 
decorrelation between Mueller--Navelet jets~\cite{Abachi:1996et}. At that time 
only the LO BFKL equation was available and predictions based on it 
failed to describe the data since it estimates too much decorrelation. 
Meanwhile, an exact fixed NLO $(\alpha_s^3)$ Monte Carlo calculation using 
the program JETRAD~\cite{Giele:1993dj} underestimated the decorrelation. In 
contrast, the Monte Carlo program HERWIG~\cite{Marchesini:1991ch} was in 
perfect agreement with the data.

\begin{figure}[htbp]
  \centering
  \includegraphics[width=8cm]{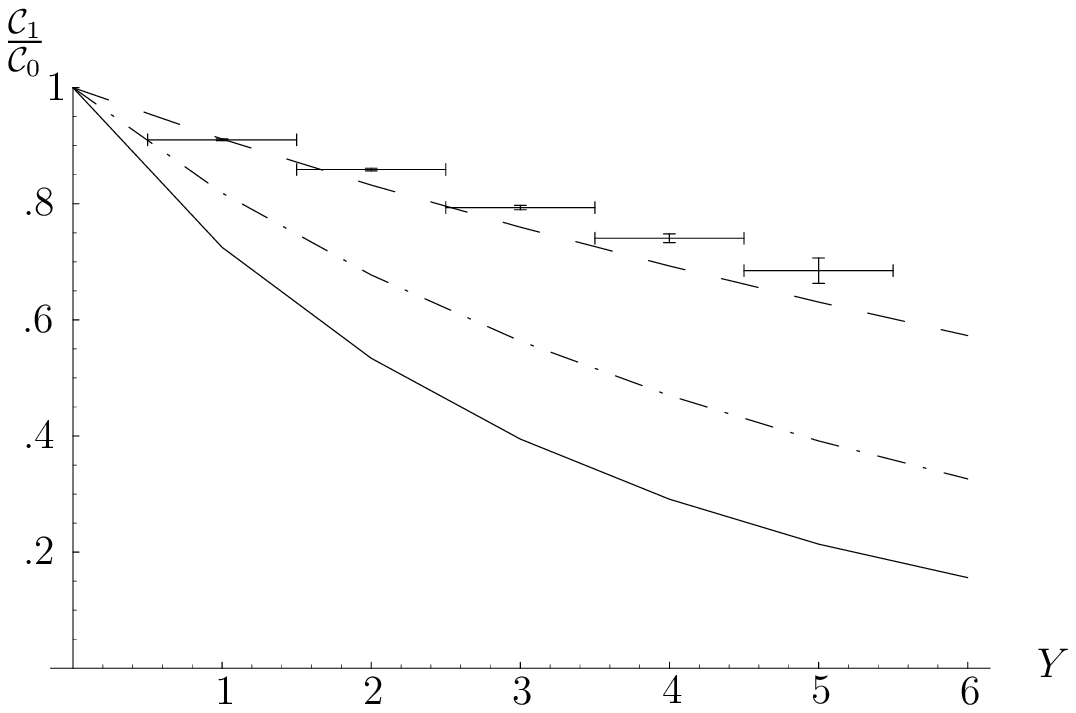}\\
  \includegraphics[width=8cm]{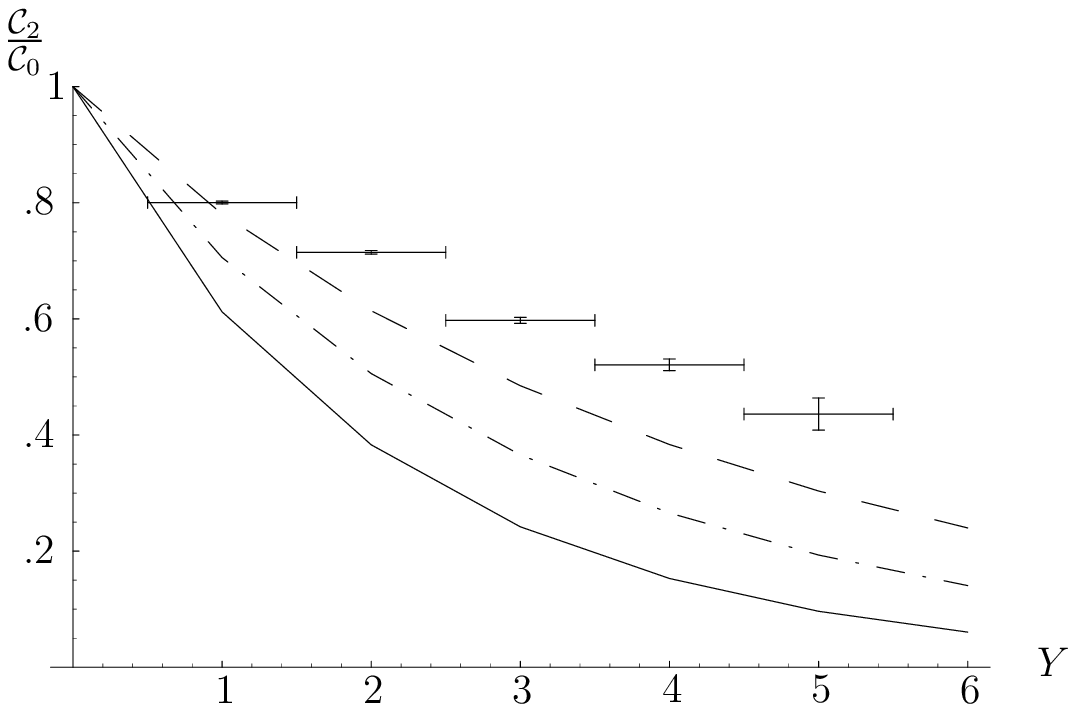}
\caption{$\left<\cos\phi\right> = {\cal C}_1/{\cal C}_0$ 
and $\left<\cos 2\phi\right> = {\cal C}_2/{\cal C}_0$ at a 
$p\bar{p}$ collider with $\sqrt{s}$ = 1.8 TeV for BFKL at LO (solid) and 
NLO (dashed). The results from the resummation presented in the text are 
shown as well (dash--dotted). Tevatron data points are taken from 
Ref.~\cite{Kim:1996dg}.}
\label{fig:tevatron1}
\end{figure}

\begin{figure}[htbp]
  \centering
  \includegraphics[width=8cm]{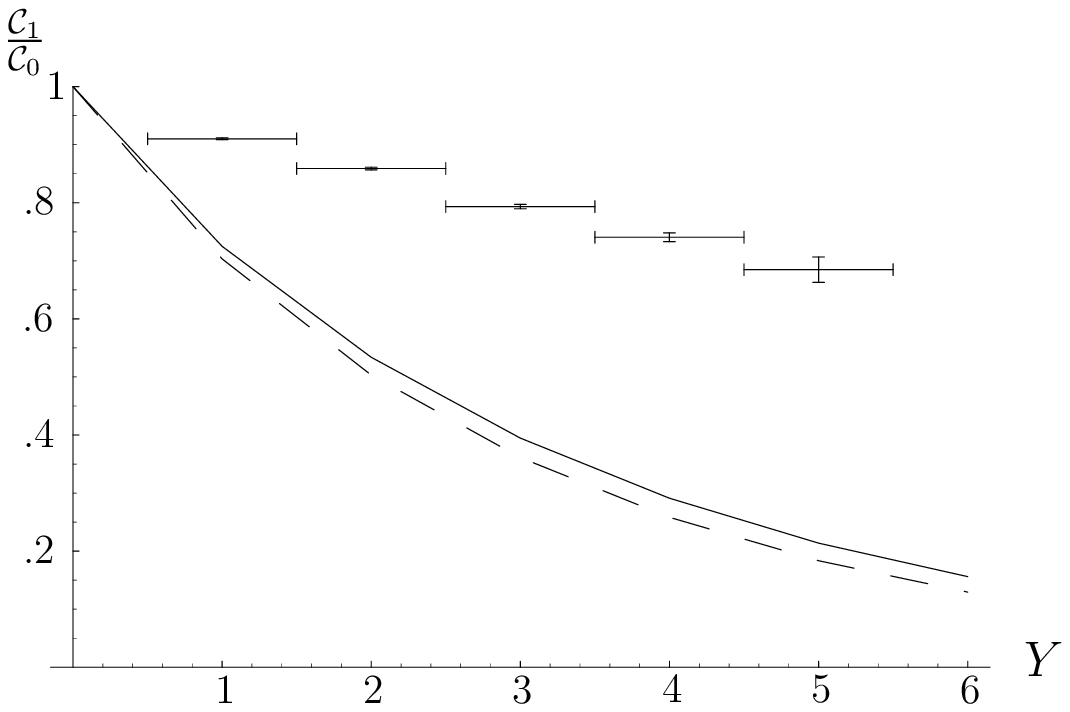}\hspace{.5cm}
  \includegraphics[width=8cm]{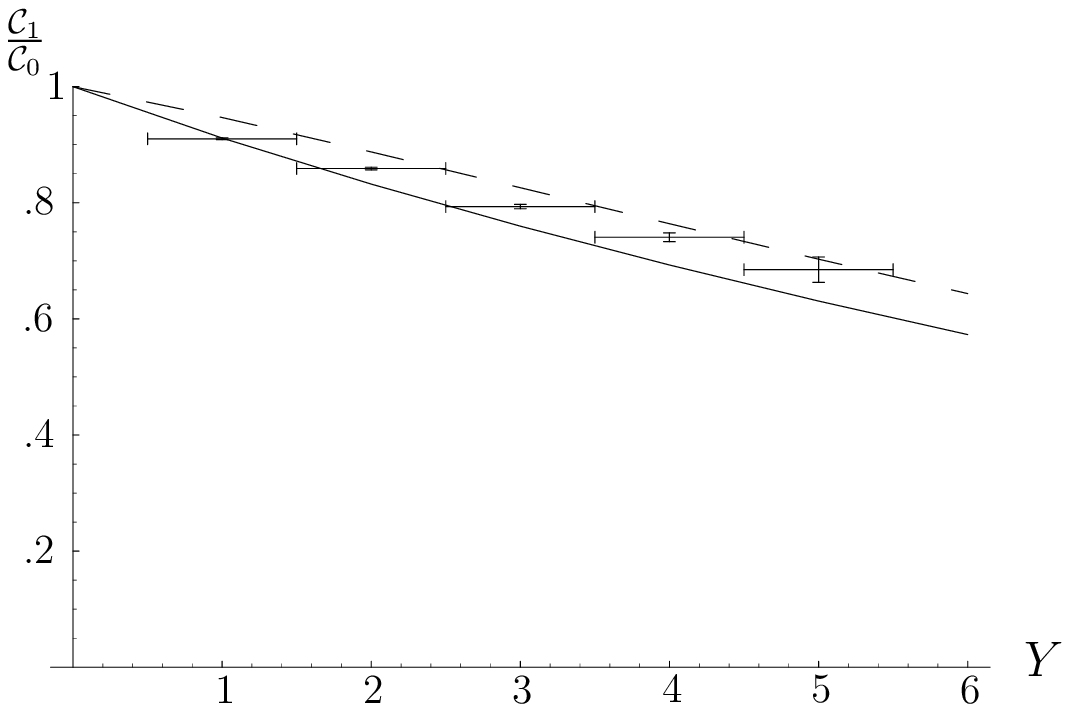}\\
  \includegraphics[width=8cm]{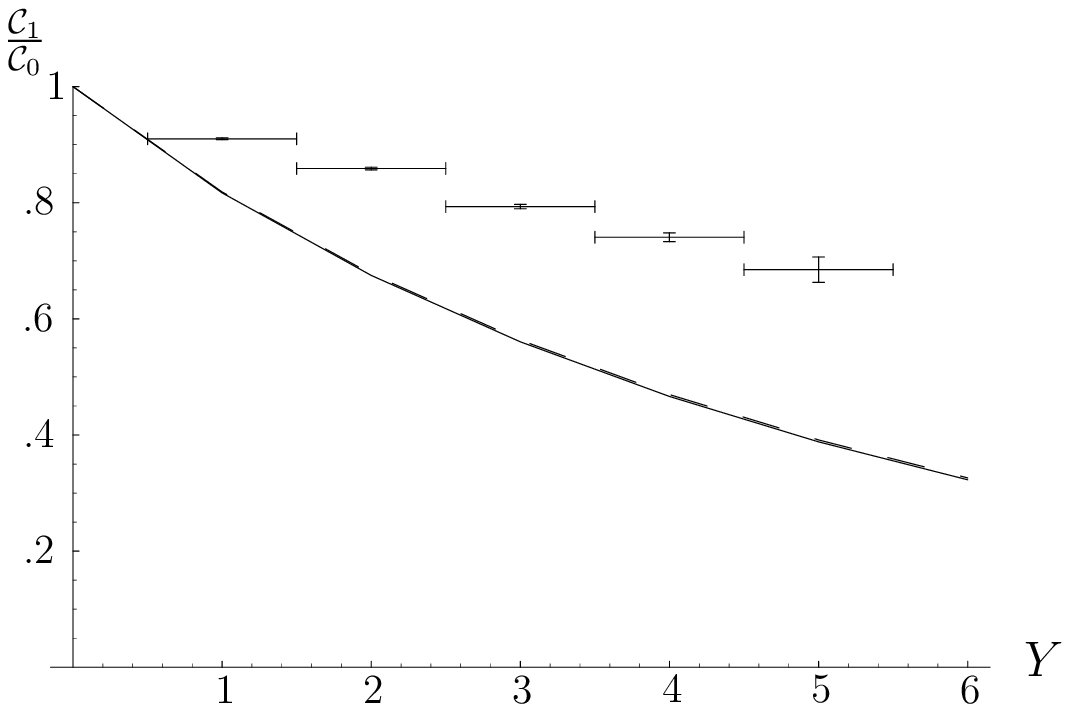}
   \caption{The same plots as in Fig.~\ref{fig:tevatron1} for 
$\left<\cos\phi\right>$ comparing the 
$\overline{\rm MS}$ renormalization scheme (solid) with the GB scheme 
(dashed). The plots correspond to LO (top), NLO (middle) and collinearly 
resummed (bottom) BFKL kernels.}
  \label{fig:tevatronresummed2}
\end{figure}

The main target of our present work is try to improve the prediction for this 
observable using the BFKL resummation introducing NLO effects in the kernel. 
We now show the effect 
of these new terms for the different observables related to the azimuthal 
angle dependence.  As mentioned before, the convenient choice of the rapidity 
variable 
${\rm Y} = \ln{\hat s / p_1 p_2}$ turns the convolution with the effective 
parton 
distributions into a simple global factor which cancels whenever we study 
ratios of cross sections or coefficients $\mathcal{C}_n$. For such 
observables, the hadronic level calculation does therefore not differ from the 
partonic one in this approximation. We start by showing in 
Fig.~\ref{fig:tevatron1} 
the Tevatron data for the average of the azimuthal angle between the 
two tagged Mueller--Navelet jets, 
$\left<\cos\phi\right> = {\cal C}_1/{\cal C}_0$ and 
$\left<\cos 2\phi\right> = {\cal C}_2/{\cal C}_0$, and compare them with 
our resummed prediction developed in the previous section using 
Eq.~(\ref{cosmphi}), which evaluates the angular mean values in terms of the 
coefficients $\mathcal{C}_n$ in Eq.~(\ref{Cn}). For comparison we also show 
the LO and standard NLO BFKL results without any further resummation of higher 
order terms. As a general trend a decrease of the amount of correlation as Y 
gets larger is obtained, and it can be seen that the NLO corrections to the 
BFKL kernel change the LO results significantly. 
For the particular cuts at the Tevatron, where the transverse momentum for one 
jet is 20 GeV and for the other 50 GeV, it turns out that the NLO calculation  
in the $\overline{\rm MS}$--scheme provides the best fit to the data. However, 
this prediction is very instable under a change of renormalization scheme 
and we cannot trust it. A first hint of this point is that if we change from 
$\overline{\rm MS}$ to GB scheme we notice that the NLO result varies more 
than the LO one. Meanwhile, the resummed prediction does not change. This 
can be clearly seen in Fig.~\ref{fig:tevatronresummed2} where we have 
calculated 
$\left<\cos\phi\right>$ in both renormalization schemes.

\begin{figure}[htbp]
  \centering
  \includegraphics[width=8cm]{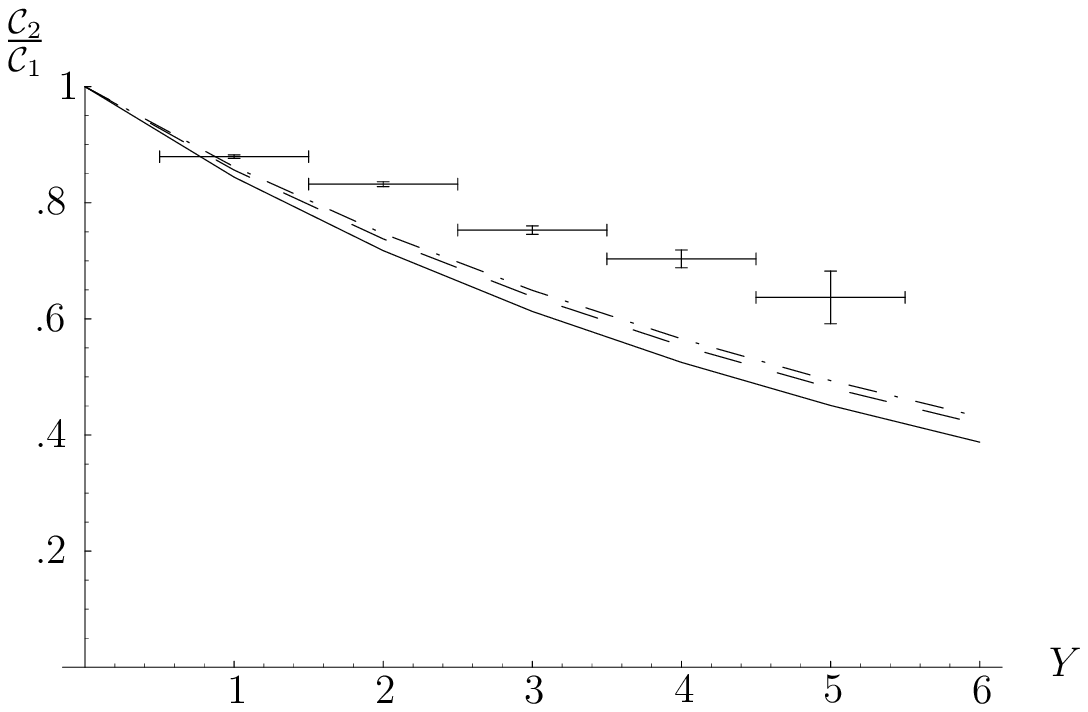}
   \caption{$\frac{<\cos2\phi>}{<\cos\phi>} = \frac{{\cal C}_2}{{\cal C}_1}$ 
as obtained from Fig.~\ref{fig:tevatron1} with LO (solid), 
NLO (dashed) and collinearly resummed (dash-dotted) BFKL kernels.}
  \label{fig:tevatron30}
\end{figure}

It is important to indicate that the convergence of our observables is poor 
whenever the coefficient associated to zero conformal spin, ${\cal C}_0$, is 
involved. If 
we eliminate this coefficient by calculating the ratios defined in 
Eq.~(\ref{Ratiosformula}) then the only dependence is on the higher $n$'s and 
the predictions are very stable under the introduction of higher order 
corrections. This is illustrated in Fig.~\ref{fig:tevatron30} where we can 
observe that the predictions at LO, NLO and with a resummed kernel for 
$\frac{<\cos2\phi>}{<\cos\phi>} = \frac{{\cal C}_2}{{\cal C}_1}$ are very 
similar. 

\begin{figure}[htbp]
  \centering
  \includegraphics[width=9cm]{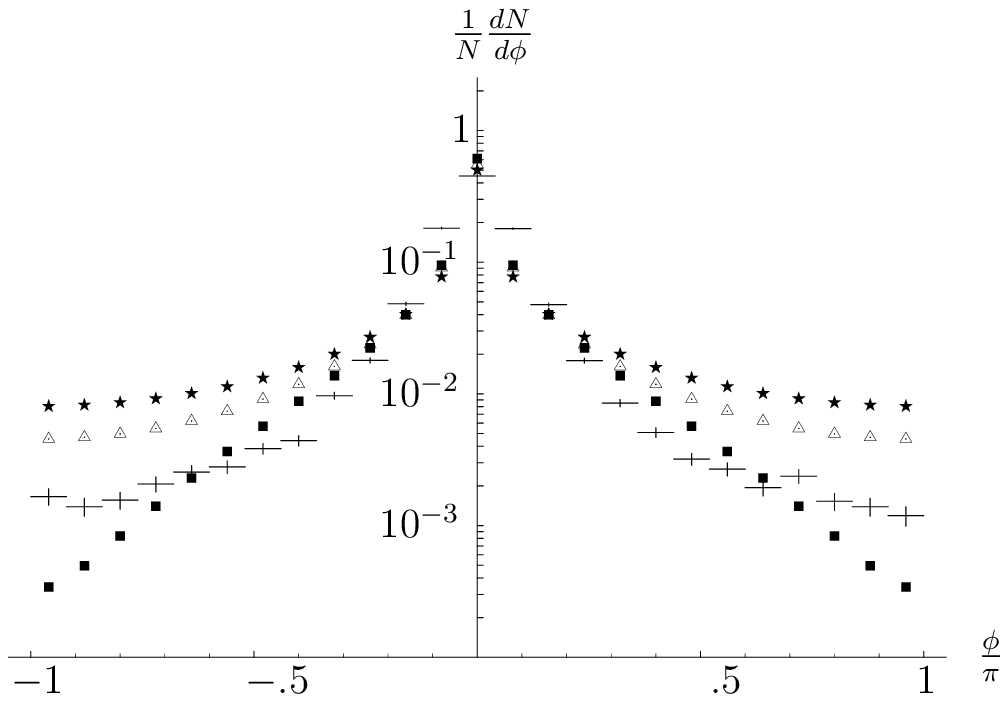}
  \includegraphics[width=9cm]{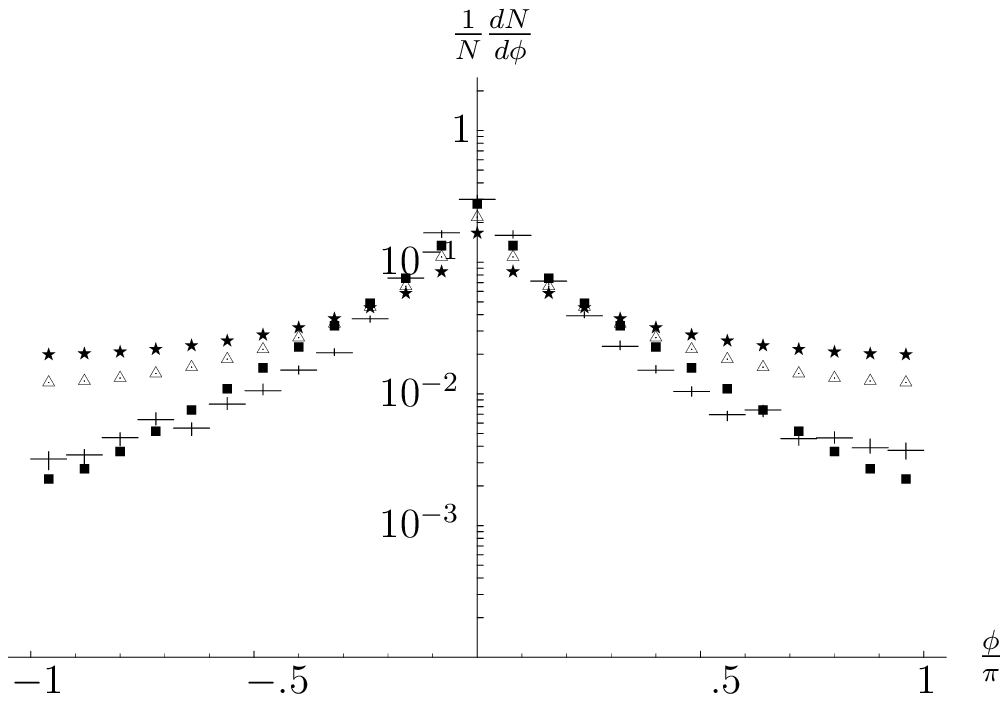}
  \includegraphics[width=9cm]{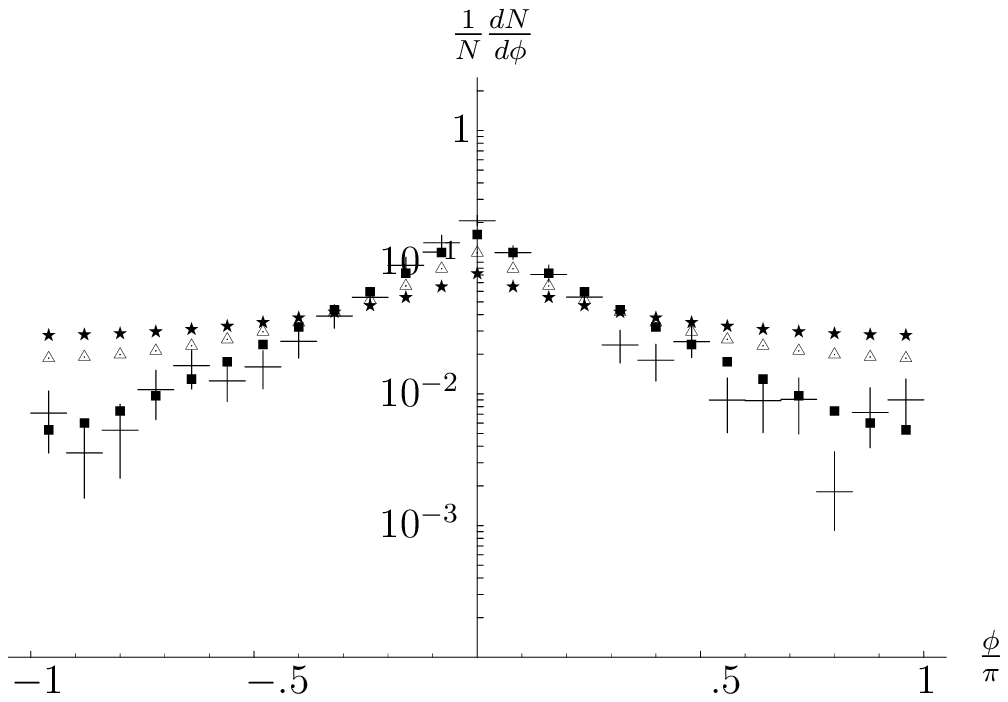}
   \caption{$\frac{1}{N}\frac{dN}{d\phi}$ in a $p\bar{p}$ collider at $\sqrt{s}$=1.8 TeV using a LO (stars), NLO (squares) and resummed (triangles) BFKL kernel. Plots are shown for Y = 1 (top), Y = 3 (middle) and Y = 5 (bottom).  Tevatron data points are taken from Ref.~\cite{Abachi:1996et}.}
  \label{fig:tevatrondsigma}
\end{figure}

We have also studied the full angular dependence by investigating the 
differential angular distribution as given in Eq.~(\ref{fullangular}). 
The {D$\emptyset$} collaboration published their measurement of this 
normalized angular distribution for different rapidity differences in 
Ref.~\cite{Abachi:1996et}. In Fig.~\ref{fig:tevatrondsigma} we compare this 
measurement with the predictions obtained in our approach using a LO, NLO, 
and resummed BFKL kernel. 
This comparison is very useful to further justify the need of a collinear 
resummation to all orders. The NLO result here presented is again in the 
$\overline{\rm MS}$--scheme, when we switch to the GB--scheme the plot 
completely 
changes becoming even negative as we approach $\phi \sim \pm \pi$. This is not 
the case in the collinearly improved calculation. We can also see that the fit to the data 
in the resummed case is much better than at LO and we have checked that the 
analysis of $\chi^2/$n.d.f. for the resummed kernel improves for larger 
rapidities. Although for the low rapidities measured at the Tevatron our 
calculation is not close to the data, the fact that 
the shape of the distribution is the correct one is very reassuring. It would 
be very interesting to have measurements of this observable at the future 
LHC at CERN where a much larger 
center--of--mass energy is accessible to investigate if a BFKL--based analysis 
fits the data better for larger rapidity differences. This available rapidity 
range is restricted rather by the geometry of the detector than by the 
energy of the colliding particles and by placing calorimeters far enough 
in the forward and backward regions it would be possible to reach about 
$10 - 12$ units of Y difference which would be very useful to gauge the 
importance of multi--Regge--kinematics. 

If large values of Y were accessible in 
the data it would be very interesting to propose other observables where BFKL 
effects should be visible. As an example, with $ \sim 12$ units of 
rapidity between the most forward and most backward jets we could tag another 
jet in the central region of the detector with the condition that the 
three jets had similar transverse momenta of $\sim 10$ GeV. The two 
regions at Y $\sim 6$ from the central jet would 
be enough for BFKL evolution. Studies of the growth with energy of this 
configuration together with double differential cross sections in the 
relative azimuthal angles between the three jets would help disentangle the 
underlying BFKL dynamics.

\begin{figure}[htbp]
\centering
  \includegraphics[width=6cm]{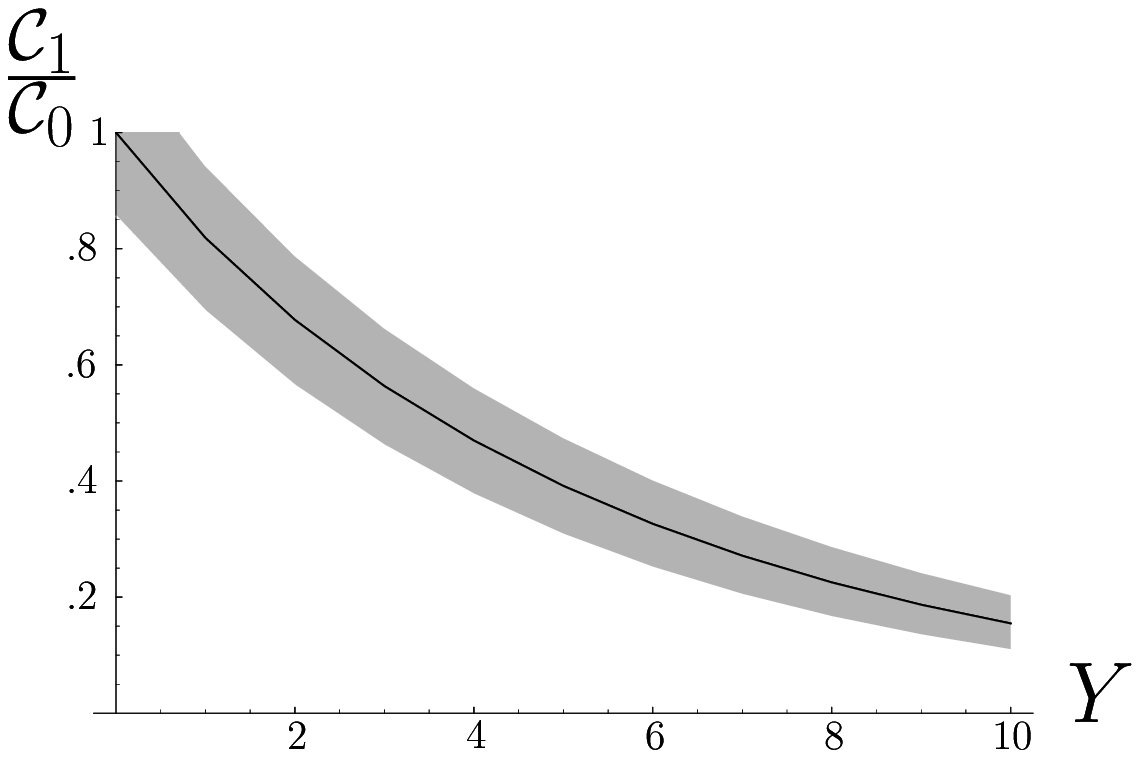}\includegraphics[width=6cm]{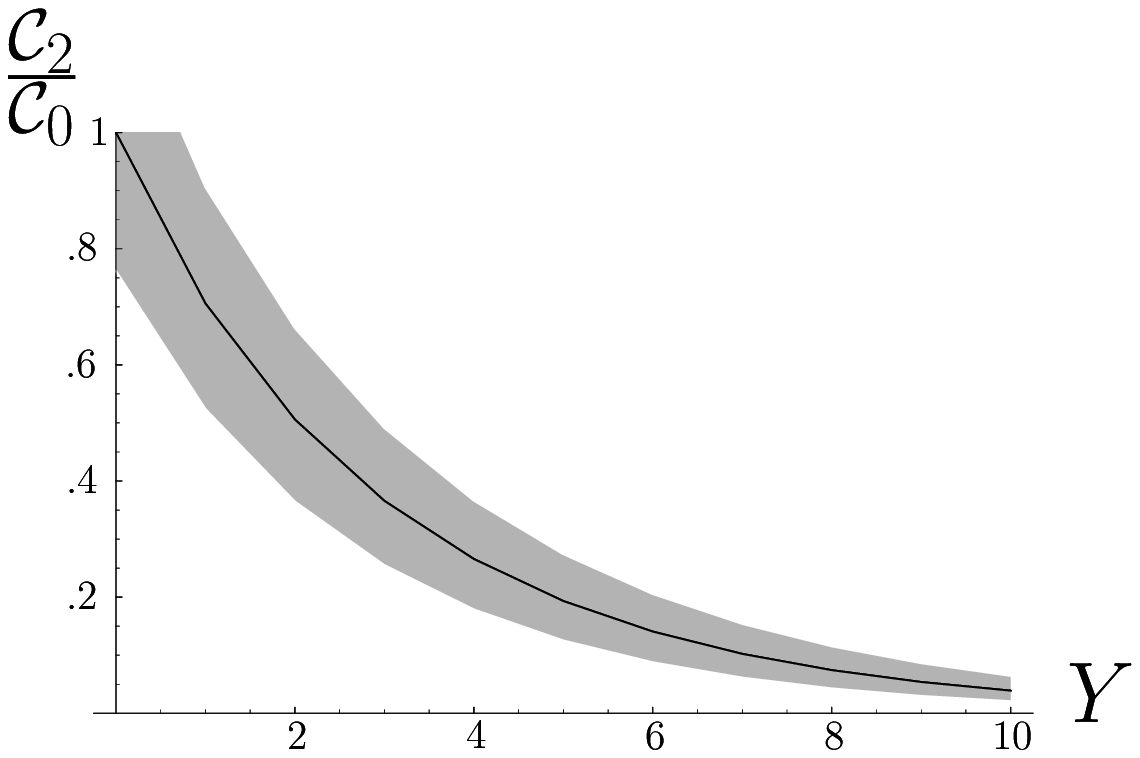}\\
\includegraphics[width=6cm]{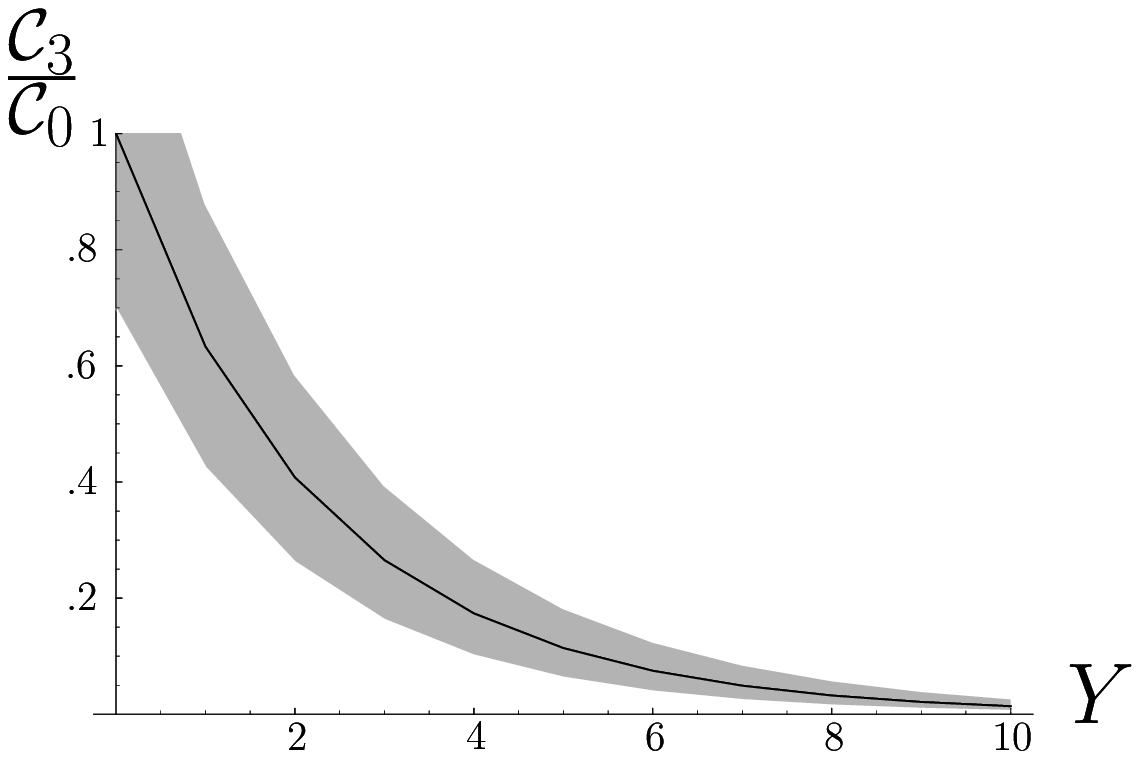} \includegraphics[width=6cm]{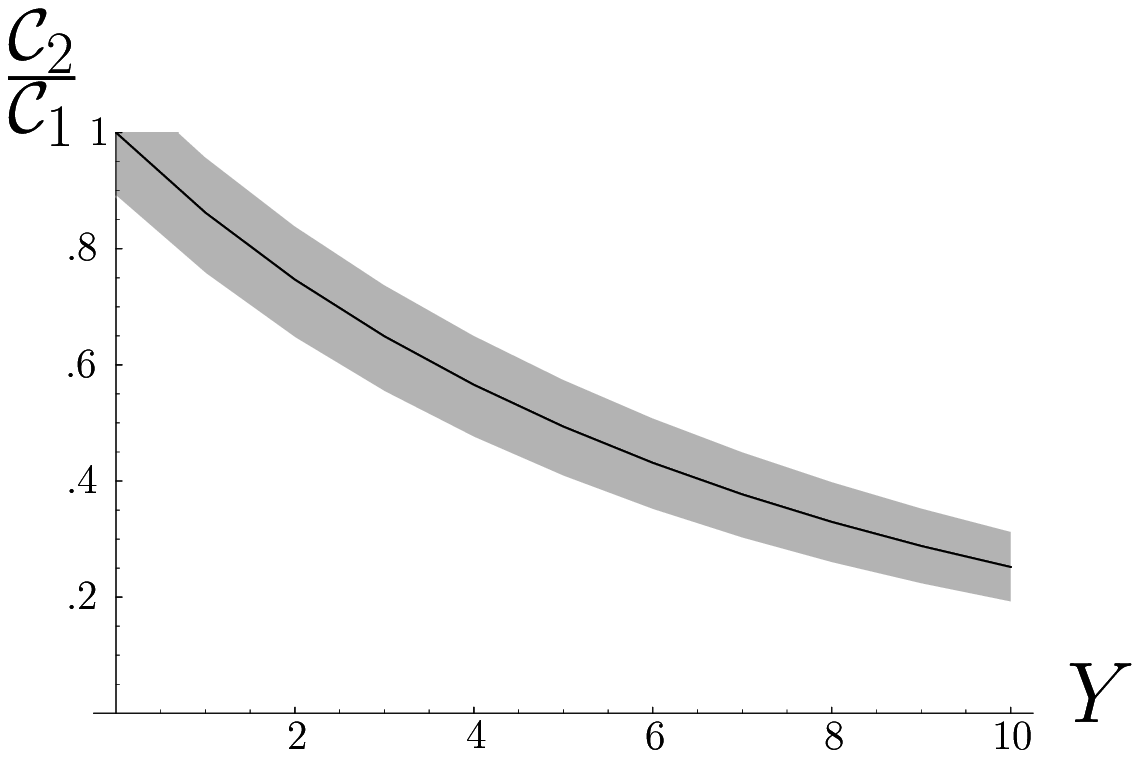}\\
\includegraphics[width=6cm]{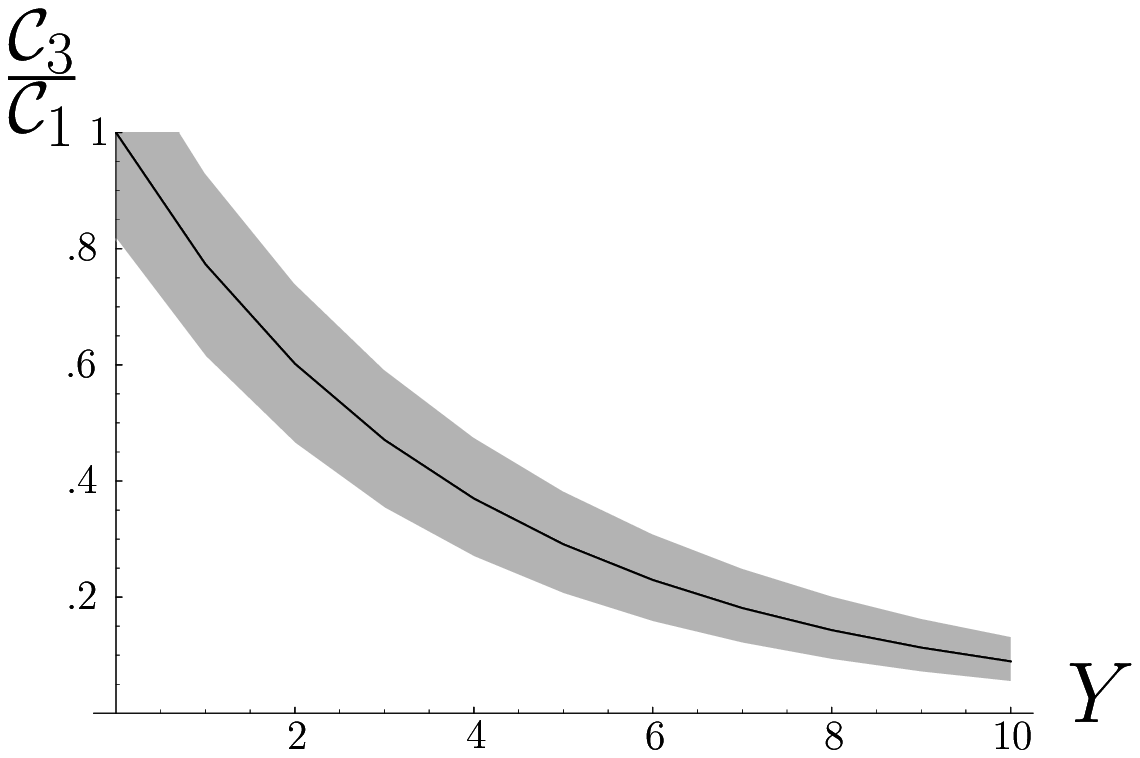}\includegraphics[width=6cm]{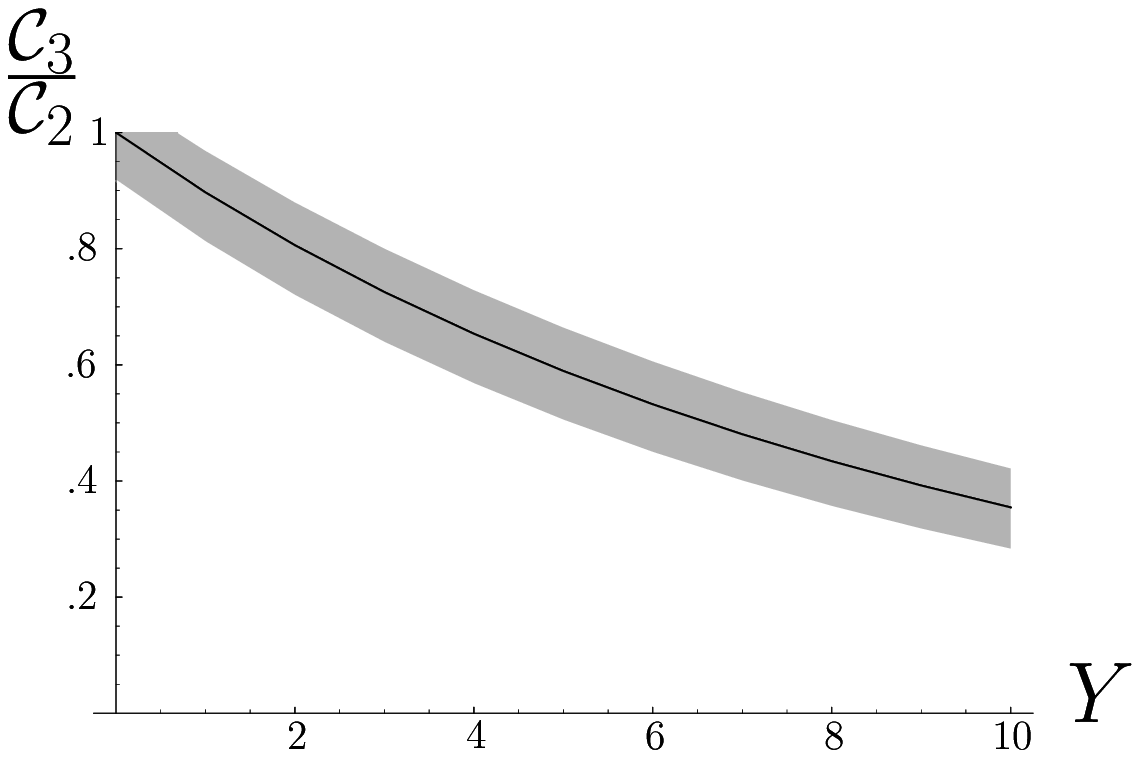} 
  \caption{Different ratios of the coefficients $\mathcal{C}_n$ obtained 
using a collinearly resummed BFKL kernel. The gray band reflects the 
uncertainty in $s_0$ and in the renormalization scale $\mu$.}
  \label{fig:lhc}
\end{figure}

\begin{figure}[htbp]
  \centering
  \includegraphics[width=9cm]{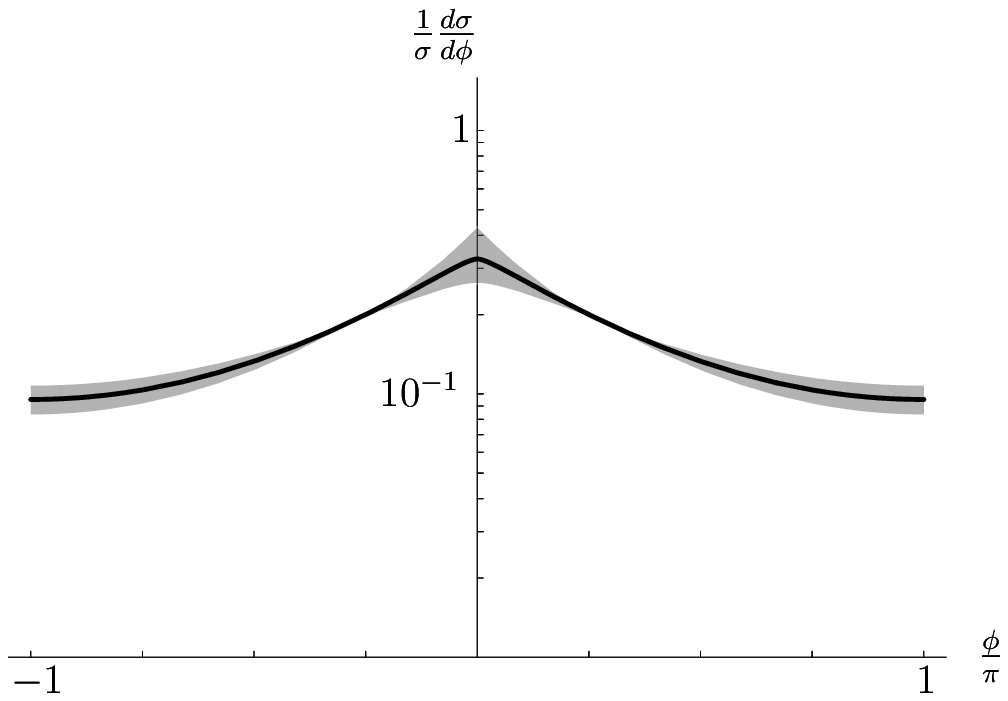}
  \includegraphics[width=9cm]{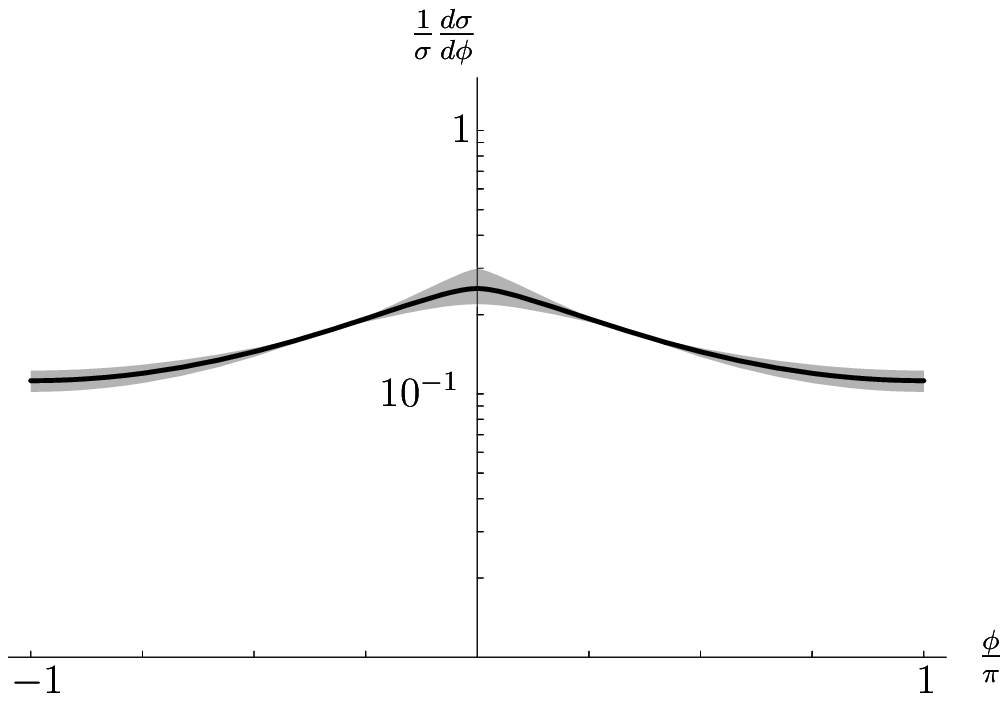}
  \includegraphics[width=9cm]{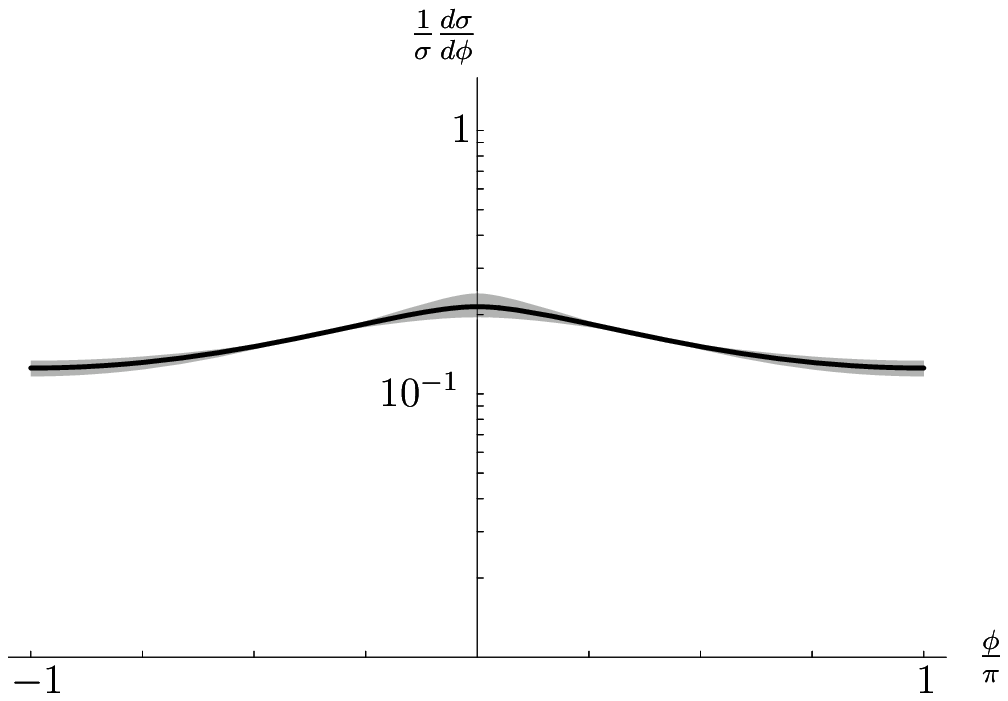}
   \caption{$\frac{1}{\sigma}\frac{d\sigma}{d\phi}$ in our resummation scheme for 
rapidities Y = 7, 9, 11 from top to bottom. The gray band reflects the 
uncertainty in $s_0$ and in the renormalization scale $\mu$.}
  \label{fig:lhcdsigma}
\end{figure}

We present numerical estimates for the ratios 
$\mathcal{C}_{m}/\mathcal{C}_{n}$ for a broader range of rapidity as 
predictions for the LHC in Fig.~\ref{fig:lhc}. 
For the differential cross section we also provide results at large Y in  
Fig.~\ref{fig:lhcdsigma}. Our calculation is not exact and we partially 
estimated the uncertainty associated to the running of the coupling and 
to the sector of the NLO Mueller--Navelet jet vertex originating from the 
splitting functions, which can be easily read off from 
Refs.~\cite{impactfactors}. It turns out that the effect on 
the overall normalization can be large, as it has been shown in 
Ref.~\cite{Kepka:2006xe}, but the influence in the ratios that we consider 
here is only of the order of a few percent. We also estimate the uncertainty 
in our choice of rapidity variable Y by varying our choice of Regge scale 
$s_0 = p_1 p_2$ by a factor of 2. We also varied the renormalization scale 
$\mu$ by the same factor, and represented the uncertainty stemming from 
these two sources by gray bands in  Fig.~\ref{fig:lhc} and 
Fig.~\ref{fig:lhcdsigma}. 
 
\section{Conclusions}

We have presented a detailed analytic study of the effects of NLO corrections 
to the BFKL kernel in the description of azimuthal angle decorrelations for 
Mueller--Navelet jets in hadron colliders. From the technical point of 
view we found that the intercepts for conformal spins other than zero have 
good convergence properties and are not largely modified when a collinear 
resummation is performed. The zero conformal spin component does need of 
this resummation to get stable results. It is possible to reduce the 
uncertainties present in our analytic study by using Monte Carlo 
techniques~\cite{MCnumerical} and work is in progress in this 
direction~\cite{SabioVera:2006rp}.

From the phenomenological side 
we have performed comparisons to the only available data related to this 
observable, which were extracted at the Tevatron many years ago. 
As the rapidity 
differences between the two tagged jets in the D$\emptyset$ experiment were 
not very large it is natural to think that a fix order calculation would do a 
better job in fitting the data than a BFKL resummation. Indeed we find that 
our results improve with respect to the LO BFKL predictions but still show too 
much azimuthal angle decorrelation. We would like to encourage the 
experimental study of this observable at the LHC at CERN with rapidity 
differences quite larger than the presently available in the literature. This 
would be very useful to investigate the importance of BFKL effects in 
multijet production in a hadronic environment. If the available 
rapidity range is large enough there is a good opportunity to study other 
jet topologies which could be dominated by multi--Regge kinematics. For these 
studies the NLO production vertex and techniques developed in 
Ref.~\cite{FSthesis,Bartels:2006hg} will be very useful.

\begin{flushleft}
{\bf \large Acknowledgments}
\end{flushleft}
We would like to thank P.~Aurenche, J.~Bartels, M.~Ciafaloni, M.~Fontannaz, H.~Jung, L.~Lipatov and L.~Motyka for very interesting discussions.  F.S. is supported by the Graduiertenkolleg 
``Zuk\"unftige Entwicklungen in der Teilchenphysik''.

\end{document}